\documentclass[a4paper,fleqn,usenatbib]{mnras}
\usepackage{newtxtext,newtxmath}
\usepackage[T1]{fontenc}
\usepackage{ae,aecompl}
\usepackage{xspace}

\usepackage{graphicx}	
\usepackage{amsmath}	
\usepackage{amssymb}	

\newcommand{\degree}{\ensuremath{^{\circ}}\xspace}

\DeclareRobustCommand{\TUSSEN}[3]{#2}

\title[RFI excision effects in 21-cm power spectra]{The impact of interference excision on 21-cm Epoch of Reionization power spectrum analyses}

\author[A. R. Offringa et al.]{
A. R. Offringa,$^{1,2}$\thanks{E-mail: offringa@astron.nl}
F. Mertens$^{2}$ and
L. V. E. Koopmans$^{2}$
\\
$^{1}$Netherlands Institute for Radio Astronomy (ASTRON), 7991 PD Dwingeloo, The Netherlands\\
$^{2}$Kapteyn Astronomical Institute, University of Groningen, PO Box 800, NL-9700 AV Groningen, the Netherlands
}

\date{Accepted 2019 January 11. Received 2018 November 9; in original form 2018 July 23}

\pubyear{2019}

\begin{document}
\label{firstpage}
\pagerange{\pageref{firstpage}--\pageref{lastpage}}
\maketitle

\begin{abstract}
We investigate the implications of interference detection for experiments that are pursuing a detection of the redshifted 21-cm signals from the Epoch of Reionization. Interference detection causes samples to be sporadically flagged and rejected. As a necessity to reduce the data volume, flagged samples are typically (implicitly) interpolated during time or frequency averaging or uv-gridding. This so-far unexplored systematic biases the 21-cm power spectrum, and it is important to understand this bias for current 21-cm experiments as well as the upcoming SKA Epoch of Reionization experiment. We analyse simulated data using power spectrum analysis and Gaussian process regression. We find that the combination of flagging and averaging causes tiny spectral fluctuations, resulting in ``flagging excess power''. This excess power does not substantially average down over time and, without extra mitigation techniques, can exceed the power of realistic models of the 21-cm reionization signals in LOFAR observations. We mitigate the bias by i) implementing a novel way to average data using a Gaussian-weighted interpolation scheme; ii) using unitary instead of inverse-variance weighting of visibilities; and iii) using low-resolution forward modelling of the data. After these modifications, which have been integrated in the LOFAR EoR processing pipeline, the excess power reduces by approximately three orders of magnitude, and is no longer preventing a detection of the 21-cm signals.
\end{abstract}




\section{Introduction}
The Epoch of Reionization (EoR) is a phase in the evolution of our Universe of which, at present, relatively little is known. A promising way to study the EoR is by using a low-frequency interferometric array to statistically detect the redshifted 21-cm signals of neutral hydrogen from the Epoch of Reionization using 21-cm power spectrum analyses \citep{liev-2002, morales-2005, furlanetto-2006, mcquinn-2006}. Several telescopes have been designed to study the 21-cm EoR power spectrum, such as the Low-Frequency Array (LOFAR; \citealt{lofar-2013}); the Donald C. Backer Precision Array for Probing the Epoch of Reionization (PAPER; \citealt{parsons-2012-arraysensitivity}); and the Murchison Widefield Array (MWA; \citealt{beardsley-2016}), and it is one of the planned key science drivers of the Square Kilometre Array (SKA).

The faint signals from the EoR are hidden behind strong galactic and extragalactic foregrounds, which are orders of magnitude brighter \citep{jelic-lofar-foregrounds-2008, bernardi-wsrt-foregrounds-II-2010}. There are several methods that are pursued to achieve a detection. First of all, a large part of the foregrounds can be subtracted from the data by creating accurate sky models  \citep{yatawatta-2013, carroll-2016, procopio-2017}. Furthermore, the foregrounds and 21-cm signals are expected to have different spectral behaviour, and are therefore distinguishable in different parts of a cylindrically-averaged power spectrum \citep{liu-2009, datta-2010-eor-foreground-subtraction, vedantham-eor-foregrounds-2012, morales-2012-eorwindow, offringa-2016}. Finally, techniques have been designed that can statistically separate the (residual) spectrally-smooth foregrounds from the spectrally-fluctuating 21-cm signals \citep{harker-2009, chapman-2013, mertens-2018}. To some level, all these techniques assume that the foregrounds are measured extremely accurately: if rapidly-fluctuating features are introduced by either the instrument or the processing that are not modelled, it may no longer be possible to separate foregrounds from 21-cm signals. This sets strong requirements on the accuracy to which the data are calibrated (\citealt{barry-2016}; \citealt{patil-2017}; \mbox{\citealt{ewall-wice-2017}}; \citealt{trott-2017}). In this paper, we analyse the contaminating effect that interference detection can introduce, assert whether these effects are strong enough to cause problems for a 21-cm power spectrum detection and introduce techniques to mitigate the issue.

All major interferometric EoR experiments use some form of radio-frequency interference (RFI) rejection, for example by detecting outlier samples and ``flagging'' these as being contaminated \citep{winkel-2006, pieflag-middelberg-2006, prasad-flagcal-2012, peck-2013, offringa-2015-mwa-rfi}. Further processing (calibration, imaging, power spectrum generation) will subsequently ignore those samples. RFI detection is most effective at high resolution \citep{offringa-2010-post-correlation-rfi-classification}, while many of the scientific goals, including EoR power spectrum studies, do not require high-resolution products. Therefore, the recorded data are typically averaged in time and frequency to a resolution of several seconds and tens of kilohertz to reduce its volume after flagging. Furthermore, pipelines that work from a gridded uv-plane or from images employ binning of visibilities based on their $uv$-values. During the gridding process, visibilities are also averaged together. If the RFI-flagged samples are removed during these data averaging steps, the irregular distribution of missing samples will result in fluctuations in the visibilities.
In this paper, we will analyse the magnitude of these effects on the 21-cm power spectrum within the context of the LOFAR EoR experiment.

In Section~\ref{sec:lofareor} we introduce the standard LOFAR EoR processing methodology. Section~\ref{sec:methods} describes the methods used in this work to analyse and mitigate the flagging excess power. In Section~\ref{sec:results} we show the impact of flagging on 21-cm power spectrum analyses, and present the results of the mitigation methods. In Section \ref{sec:conclusions}, we draw our conclusions.

\section{The LOFAR EoR processing pipeline} \label{sec:lofareor}
This paper analyses results within the context of the LOFAR EoR data processing methodology \citep{patil-2016}. Therefore, we briefly summarize the default LOFAR EoR processing steps, which are: i) RFI rejection using \textsc{aoflagger} \citep{offringa-2012-scale-invariant-rank-operator}; ii) data averaging using \textsc{dppp}; iii) (optionally) data compression using \textsc{dysco} \citep{offringa-2016-dysco}; iv) direction-independent calibration using \textsc{dppp} or \textsc{sagecal} \citep{kazemi-2011-sagecal}; v) direction-dependent compact source removal using \textsc{sagecal-co} \citep{yatawatta-co-2016}; vi) imaging using \textsc{wsclean} \citep{offringa-wsclean-2014} or \textsc{excon} \citep{yatawatta-2014-excon}; vii) residual foreground removal using GPR \citep{mertens-2018}; and viii) power spectrum calculation using both a Python and a C++ power spectrum pipeline (\S \ref{sec:power-spectrum-pipelines}).

Typical LOFAR EoR observations are performed at a time and frequency resolution of 2\,s and 3\,kHz, respectively, which will be referred to as ``high resolution data'' in the paper. Because RFI detection is most effective at high time and frequency resolution \citep{offringa-2010-post-correlation-rfi-classification}, it is the first processing step that is performed after recording the data. Afterwards, the data is averaged down by about a factor of 100 to typical resolutions of 10~s and 40 to 60~kHz. This data product will be referred to as the ``low resolution data''. Averaging decreases the data volume considerably, while it is still of high enough resolution to not cause any significant time or frequency smearing within the 5\degree LOFAR primary beam.

\begin{table}
\caption{Details of simulated dataset \& observed RFI flags.} 
\label{tbl:observation}
\begin{center}
\begin{tabular}{|l|r|}
\hline
\hline
Observing start & 4 Feb 2018, 16:49 (UTC)\\
Observing end   & 5 Feb 2018, 6:43 (UTC) \\
LOFAR ID & L628584 \\
Duration & 50001\,s ($\sim$ 14\,h) \\
\hline
RA       & 0h 00 \\
Dec      & 90\degree00 \\
\hline
Band & 113.8--128.1\,MHz \\
21-cm redshift & 11.5--10.1 \\
\hline
RFI percentage & 1.12\,\% \\
Polarization & XX, XY, YX, YY \\
\hline
\hline
\end{tabular}
\end{center}

\end{table}

\section{Simulated data \& Methods} \label{sec:methods}

\begin{figure}
	\includegraphics[width=\columnwidth]{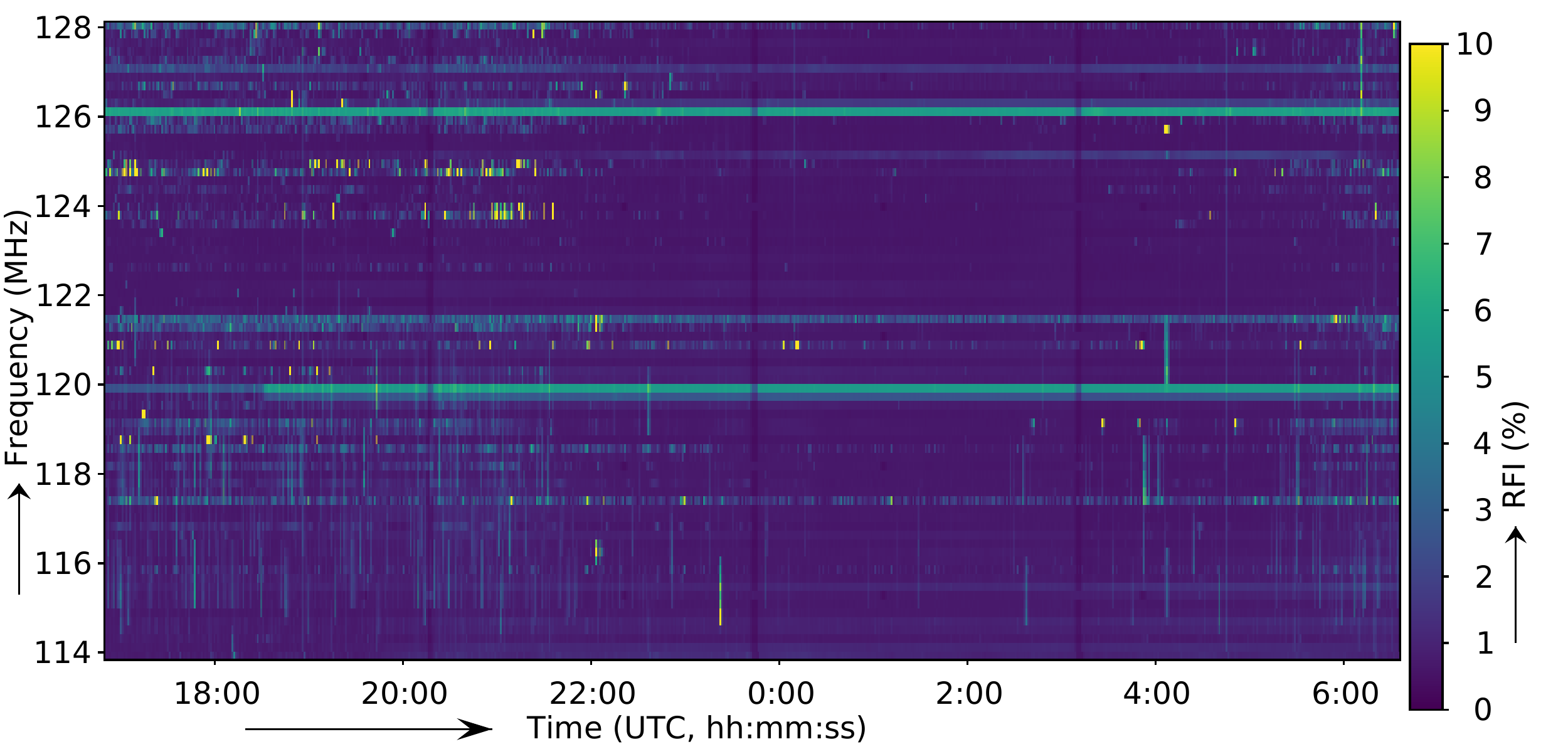}%
	\caption{Detected RFI occupancy over time and frequency. The RFI detection is performed at a resolution of 3 kHz / 2 s. The occupancy shown here is averaged over all baselines in bins of 200 kHz / 12 s. }
	\label{fig:flagging-occupancy-spectrum}
\end{figure}

To accurately simulate the effect of flagging, we use the RFI detection flags from an actual LOFAR observation of the North Celestial Pole (NCP) --- one of the target fields of the LOFAR EoR project. During typical processing, the flagged high-resolution data with a resolution of 2\,s and 3\,kHz are not written to disk, but are immediately averaged down, and the high-resolution input data are removed afterwards to free up space for further observations. To gain access to the high resolution flags for this study, the full raw data are intercepted before averaging down and are written to disk. Subsequently, \textsc{aoflagger} is run on the data without averaging afterwards. Thereafter, the observed visibilities are replaced by simulated visibilities for a realistic model. The result is a simulated high-resolution visibility set with realistic RFI detection flags.

Table~\ref{tbl:observation} summarizes the observation that was used for RFI detection. \citet{offringa-lofar-environment-2013} have previously studied the RFI environment of LOFAR using two 24 hour observations targetting the NCP. They show that the 114-128 MHz frequency range is one of the least RFI affected frequency ranges in the HBA, which is also reflected in the detected RFI occupancies (1.1\% found in this study vs. an average of 3.2\% in the HBA as found by \citealt{offringa-lofar-environment-2013}). Fig.~\ref{fig:flagging-occupancy-spectrum} shows an overview over time and frequency of the detected RFI occupancies of the data used in this work.

Although an extensive model of the NCP field is available, we chose to use a simulated point-source model based on a population study, which prevents selection effects and artefacts in the model. We use a simple randomly generated population distribution that follows the empirically determined distribution by \citet{franzen-2016}:
\begin{equation}
\frac{dN}{dS} = 6998\,S^{-1.54} \textrm{Jy}^{-1} \textrm{Sr}^{-1}.
\end{equation}
We analytically predict (using the direct Fourier transform) the contribution for sources with a flux density between 10~mJy and 10~Jy, and assign a random spectral index (SI) to each source with an average of
$\alpha=-0.8$ and a standard deviation of 0.2 (with $S(\nu)=S_0 (\nu/\nu_0)^\alpha$). This is a reasonable distribution at the corresponding frequency (e.g., \citealt{hurley-walker-2017-gleam}),
albeit that we ignore flattening of fainter (starburst) galaxies and ignore special classes of sources such as USS, CSS or GPS sources that can have steep or curved spectra at the frequencies of interest (see \citealt{callingham-2017} for an overview). We also do not simulate any diffuse emission.
Using the \textsc{dppp} software, the flux density contribution of each source in our final source model is predicted at the resolution of our data and multiplied by the corresponding gain of the LOFAR HBA beam model that combines the station array factor and the tile beam. Because of the high resolution of our data, this step is the most expensive step in the processing, and takes about a week of computing on 16 high-performance nodes, each with 40 CPU-cores.

After having predicted the simulated foregrounds into the observation, we create two averaged sets from these: one in which the flagged samples are excluded in the averaging, which is how RFI flagging would affect a regular observation, and one in which the flags are ignored and all data is used, which simulates an observation that is completely free of RFI.

\subsection{Image-based calculation of power spectra} \label{sec:power-spectrum-pipelines}
The LOFAR EoR project makes use of two independent power spectrum pipelines: a Python pipeline and a C++ pipeline. The Python pipeline is written to integrate Gaussian process regression, while the C++ pipeline is used for quick analysis. Having multiple pipelines has been very useful for the verification of results, as also noted by other teams \citep{jacobs-2016}. When using the same settings and data, our two pipelines produce similar results.

For calculating the power spectra, we follow the definitions from \citet{parsons-2012-arraysensitivity} and \citet{trott-2016-chips}, where the power spectrum is calculated as\footnote{Our definition of $P$ differs from Eq.~(2) in \citet{parsons-2012-arraysensitivity}, where a factor of $V$ is mistakenly left out. From their definitions, it can be shown that their Eq.~(3) is correct, and that equation implies the factor of $V$ in the definition of $P$.}
\begin{equation} \label{eq:powerspectrum}
 \hat{P}(\mathbf{k}) \equiv V | \tilde{T}(\mathbf{k}) |^2.
\end{equation}
Here, $V$ is the comoving volume of the data cube, $\tilde{T}$ is defined as the (normalized) discrete Fourier transform of $T$:
\begin{equation}
 \tilde{T}(\mathbf{k}) \equiv \frac{1}{N} \sum\limits_\mathbf{x} T(\mathbf{x}) e^{-i\mathbf{k}\cdot\mathbf{x} },
\end{equation}
$x$ is a physical coordinate, $k$ is inverse scale and $N$ is the number of voxels in the data cube.

For this work, both pipelines use \textsc{wsclean} \citep{offringa-wsclean-2014} as a first step to grid the data and decrease the data volume. Nevertheless, the Python pipeline can also work directly from ungridded visibilities \citep{ghosh-2018}.
We perform the imaging with increased-accuracy settings for \textsc{wsclean}, which includes a larger gridding kernel, higher oversampling rate and increased number of $w$-layers, compared to default settings of \textsc{wsclean}.
The output of the imager is a naturally-weighted primary-beam-corrected image with units of Jy/beam. Commonly, conversion to Kelvin is performed by fitting the synthesized beam to a Gaussian, followed by evaluation of
\begin{equation} \label{eq:jansky-to-kelvin}
 T(\mathbf{x}) \equiv S_\textrm{Jy/B}(\mathbf{x}) \frac{ 10^{-26} c^2 }{ 2 k_B \nu^2 \Omega_\textrm{psf}},
\end{equation}
with $S_\textrm{Jy/B}(\mathbf{x})$ the data cube in units of Jy/beam. However, we found that the synthesized beam of LOFAR deviates from a Gaussian function, causing the power to be underestimated by a factor of 1.5 when using this approach.
To overcome this, \textsc{wsclean} stores\footnote{Stored as the \texttt{WSCNORM} keyword inside the FITS file.} the factor which it has divided the data by, and from which the ``per beam'' term can be calculated. This allows us to accurately convert the image to units of Jy/pixel\footnote{It is not strictly necessary to convert from Jy/beam to Jy/pixel, because in Eq.~\ref{eq:normalized-ft} the scaling factor appears in both the numerator and denominator of the rightmost division, and therefore cancels out. However, the factor is required when propagating the errors.}.
While naturally weighted images lead to the best power spectrum sensitivity \citep{morales-2009}, it requires normalization of the $uv$-cells in order not to bias the power spectrum. We do this by dividing the $uv$-plane of the data by the $uv$-plane of the point-spread function (PSF).
In principal, this can lead to undefined values when the $uv$-plane is not fully covered. The $uv$-plane corresponding to the PSF can in this case contain very small values due to rounding errors and small gridding kernel values. Propagating the PSF $uv$ values and using these as data weights mostly solves this issue and improves the sensitivity of the final power spectrum.
Both our pipelines do this. 
In the case of LOFAR, the perpendicular scales that we target correspond with a baseline range of $50-250 \lambda$, and all $uv$-cells are sufficiently covered within this range by the LOFAR array configuration.

Combining the above, the Fourier transform $\tilde{T}$ is calculated as
\begin{equation} \label{eq:normalized-ft}
 \tilde{T}(\mathbf{k}) = \frac{ 10^{-26} c^2 }{2 k_B \Omega_A N_\nu} \sum_\nu \left( \frac{1}{\nu^2} e^{-i \nu}
 \frac{
   \sum\limits_{\textbf{l}} S(\mathbf{x}) e^{-i \mathbf{u} \cdot \textbf{l}}
 }{
   \sum\limits_{\textbf{l}} P(\mathbf{x}) e^{-i \mathbf{u} \cdot \textbf{l}}
 } \right),
\end{equation}
with $P$ the value of the PSF. Because of lost edge channels due to the poly-phase filter of LOFAR, the bandwidth is non-uniform. Moreover, as mentioned before, each data sample has an associated weight with it, which is to be used in the line-of-sight Fourier transform. Both our pipelines solve this by performing an inverse covariance weighted, least-squares spectral analysis (LSSA) of the line-of-sight Fourier transform operation \citep{trott-2016-chips}.

In simulations where the foreground has not been subtracted from the data, we use a window function to prevent leakage from the wedge to higher $k_\parallel$ values and adapt the box volume $V$ to accommodate for the decreased line-of-sight dimension, for example a Blackman-Nuttall window \citep{nuttall-1981} as described by \citet{vedantham-eor-foregrounds-2012}. This step is normally not taken when processing real data within the LOFAR EoR project: in that case, power spectra are made from direction-dependently calibrated data, and a window function is not necessary because the foreground wedge is not present.

\begin{figure*}
	\includegraphics[width=6cm]{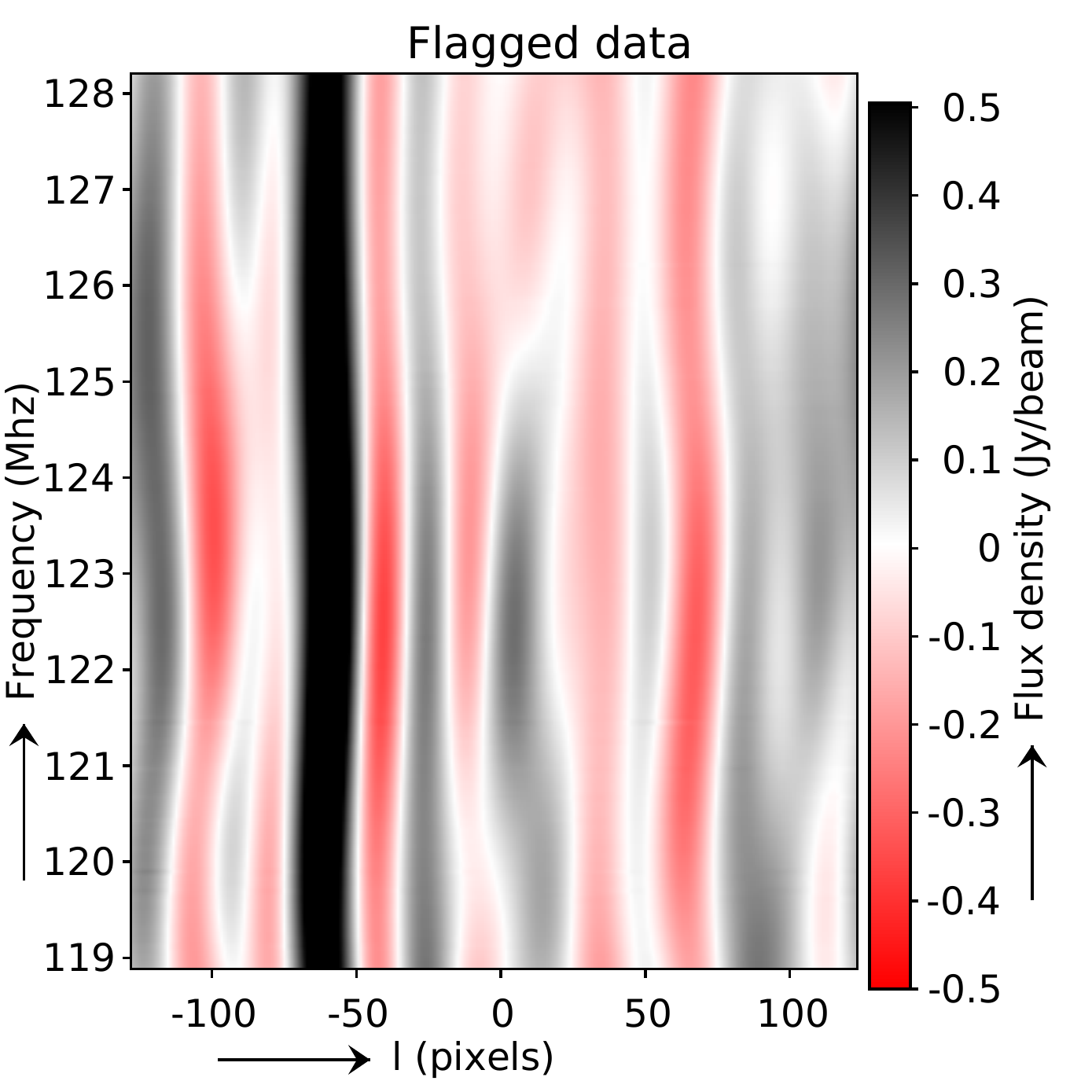}%
	\includegraphics[width=6cm]{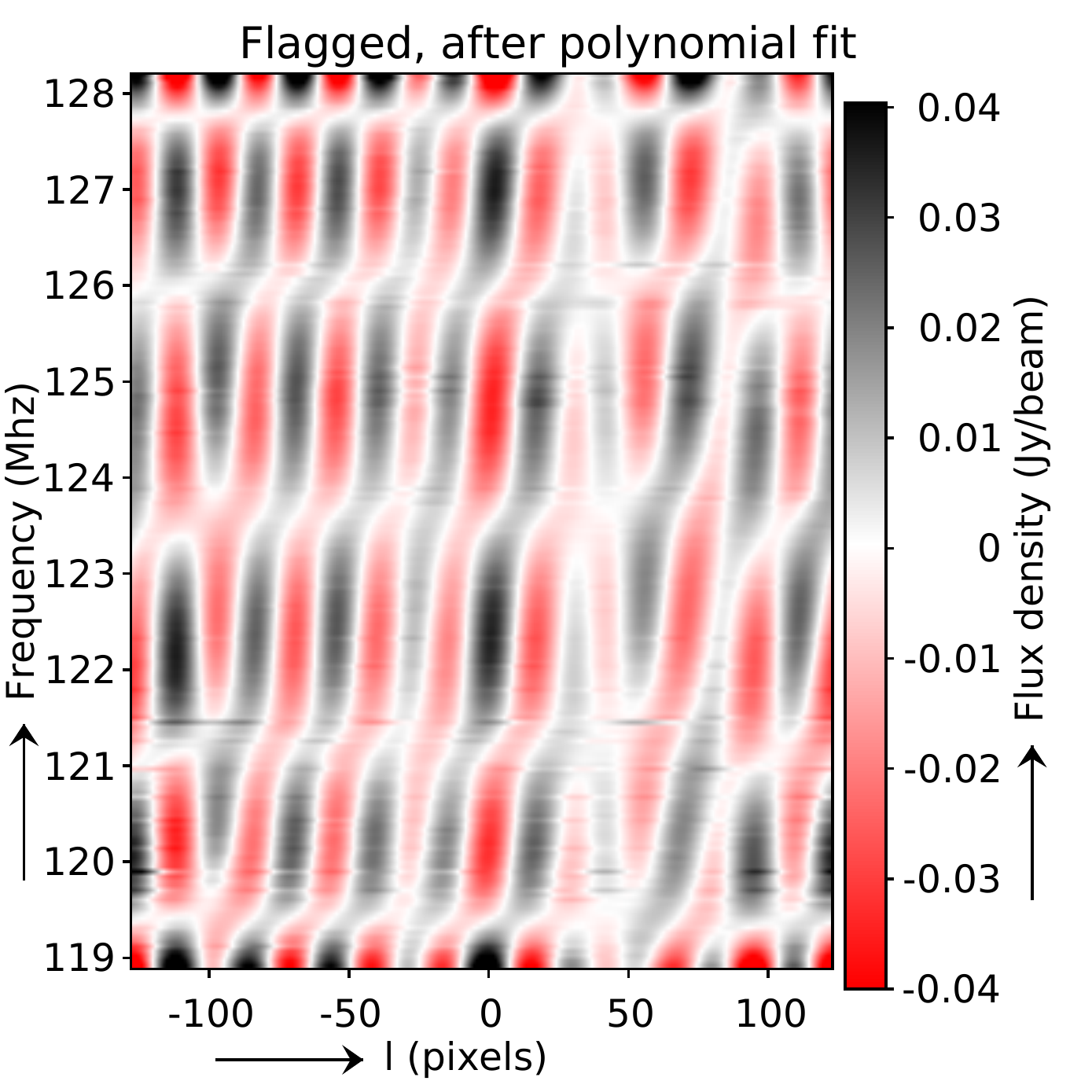}%
	\includegraphics[width=6cm]{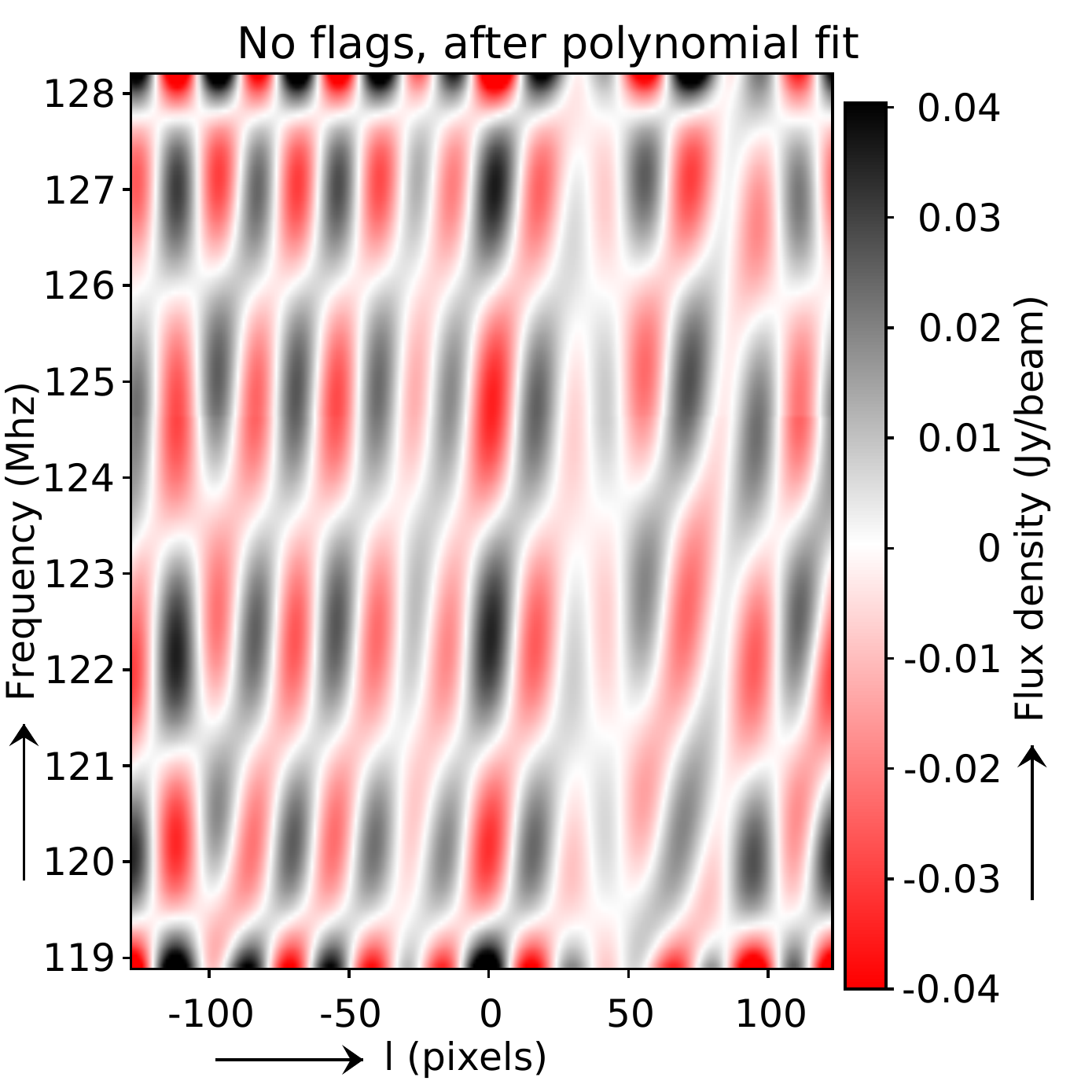}%
	\caption{Slices through a simulated image cube, showing the flux density as a function of frequency (vertical) and spatial scale (horizontal). Data that has been flagged (left image) appears to behave smoothly in frequency. However, after subtracting a 5\textsuperscript{th}-order polynomial fit, some fine-scale fluctuations are visible. These are not present when no data has been flagged (right image). }
	\label{fig:lfreq-slices}
\end{figure*}

\subsection{Gaussian Process Regression}
After calibration and direction-dependent compact source removal of data
in the LOFAR EoR project, the remaining
foregrounds, composed of extragalactic emission below the confusion noise level
and diffuse galactic emission, are still approximately 3 to 4 orders of
magnitude brighter than the 21-cm signal. The LOFAR EoR project uses the technique of Gaussian
Process Regression (GPR; \citealt{mertens-2018}) to model and remove these residual foregrounds. In
this framework, the different components of the observations, including the
astrophysical foregrounds, mode-mixing contaminants, and the 21-cm signal, are
modelled as a Gaussian process (GP). A GP is the joint distribution of a
collection of normally distributed random variables~\citep{rasmussen-2005}. The
covariance matrix of this distribution is specified by a covariance function,
which defines the covariance between pairs of observations (e.g. at different
frequencies). The covariance function determines the structure that the GP will
be able to model, for example its smoothness. In GPR, we use GP as parametrized
priors, and the Bayesian likelihood of the model is estimated by conditioning
this prior to the observations. Standard optimization or MCMC methods can be
used to determine the parameters of the covariance functions.

Formally, we model our data $\mathbf{d}$ observed at
frequencies $\mathbf{\nu}$ by a foreground, a 21-cm and a noise signal
$\mathbf{n}$~\citep{mertens-2018}:
\begin{equation}
\mathbf{d} = f_{\mathrm{fg}}(\boldsymbol{\nu}) + f_{\mathrm{21}}(\boldsymbol{\nu}) +
\mathbf{n}.
\end{equation}
The foreground signal can be separated from the 21-cm signal by exploiting their
different frequency behaviour: the 21-cm signal is expected to be largely uncorrelated on
scales of a few MHz, while the foregrounds are expected to be smooth on that
scale. The covariance function of our GP model can then be composed of a
foreground covariance function $K_{\mathrm{fg}}$ and a 21-cm signal covariance
function $K_{\mathrm{21}}$,
\begin{equation}
K = K_{\mathrm{fg}} + K_{\mathrm{21}}.
\end{equation}
The foregrounds covariance kernel itself is decomposed in two, accounting for
the large frequency coherence-scale of the intrinsic foreground and the smaller
frequency coherence-scale (about 1 to 5 MHz) of the mode-mixing component. We
use an exponential covariance function for the 21-cm signal, as we found that
it was able to match well the frequency covariance from simulated 21-cm
signal~\citep{mertens-2018}.

The joint probability density distribution of the observations $\mathbf{d}$ and
the function values $
\mathbf{f}_{\mathrm{fg}}$ of the foreground model $\mathbf{f}_{\mathrm{fg}}$ at the same
frequencies $\nu$ is then given by,
\begin{equation}
\left[ \begin{array}{c} \mathbf{d} \\ \mathbf{f_{\mathrm{fg}}} \end{array}\right] \sim  
\mathcal{N}\left( \left[\begin{array}{c} 0 \\ 0 \end{array}\right], \left[ \begin{array}{cc} (K_{\mathrm{fg}} + K_{\mathrm{21}})  + \sigma_n^2 I & K_{\mathrm{fg}} \\ K_{\mathrm{fg}} & K_{\mathrm{fg}} \end{array}  \right] \right). 
\end{equation}
using the shorthand $K \equiv K(\nu, \nu)$. After GPR, the
foregrounds part of the model is retrieved:
\begin{align}
\label{eq:gpr_predictive_mean_eor}
E(\mathbf{f}_{\mathrm{fg}}) &= K_{\mathrm{fg}}\left[K + \sigma_n^2
I\right]^{-1} \mathbf{d}\\
\label{eq:gpr_predictive_cov_eor}
\mathrm{cov}(\mathbf{f}_{\mathrm{fg}}) &= K_{\mathrm{fg}} - K_{\mathrm{fg}}\left[K + \sigma_n^2 I\right]^{-1}K_{\mathrm{fg}}.
\end{align}
and is subtracted from the original data.

\subsection{Spectral fluctuations caused by data flagging}
To introduce the symptoms of flagged data and demonstrate why a study is necessary, we compare two simulated image cubes: a flagged cube and a non-flagged ``ground-truth'' cube. For the flagged cube, the high-resolution RFI detection flags from a real LOFAR observation are transferred before averaging the observation down. Fig.~\ref{fig:lfreq-slices} shows slices through the spectral and spatial direction of several image cubes. The left plot is from data that includes flags, and shows smooth structure from sources and their sidelobes. After removing a 5\textsuperscript{th}-order polynomial fit from each line-of-sight (in $uv$ space), about an order of magnitude of flux from the data is removed, revealing residual higher order, smooth structure from the sources, but also rapid spectral fluctuations on the order of tens of mJy (Fig.~\ref{fig:lfreq-slices}, centre). These fluctuations are only present when the flags from RFI detection are added to the high-resolution data, before averaging down. The right plot of Fig.~\ref{fig:lfreq-slices} shows the simulated data without flags, after a polynomial fit.

\begin{figure*}
\includegraphics[width=8cm]{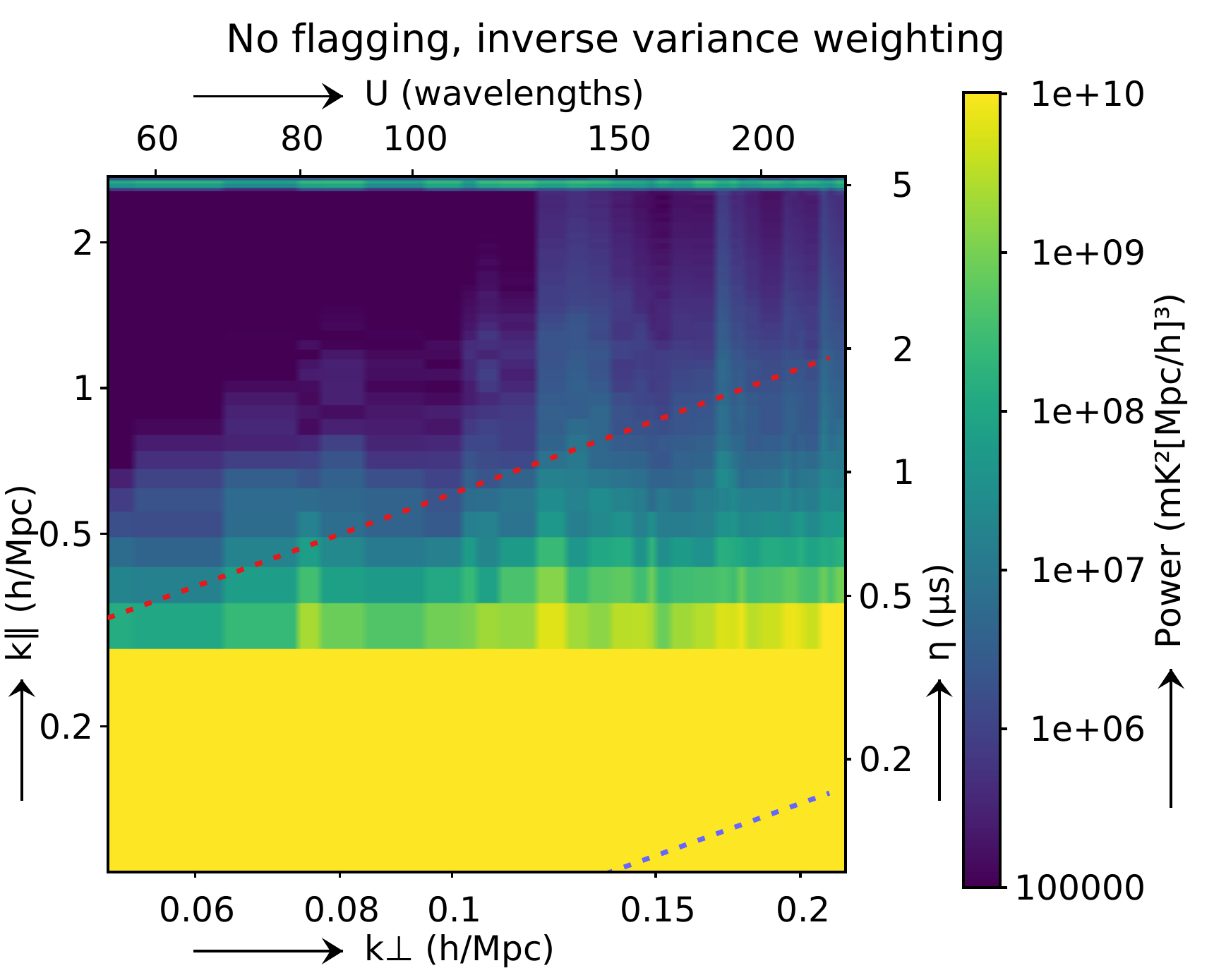}%
\includegraphics[width=8cm]{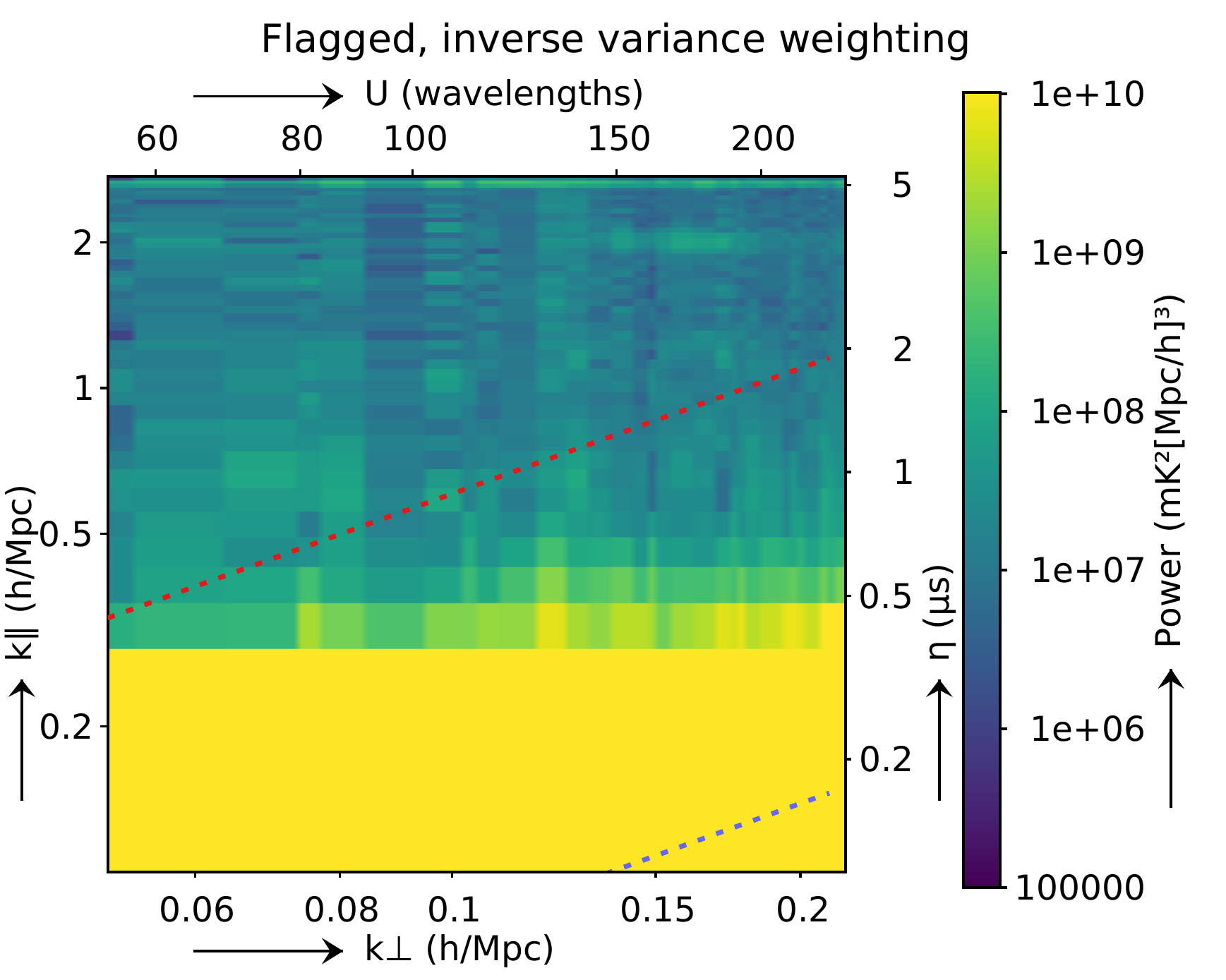}\\%
\vspace*{1mm}%
\includegraphics[width=8cm]{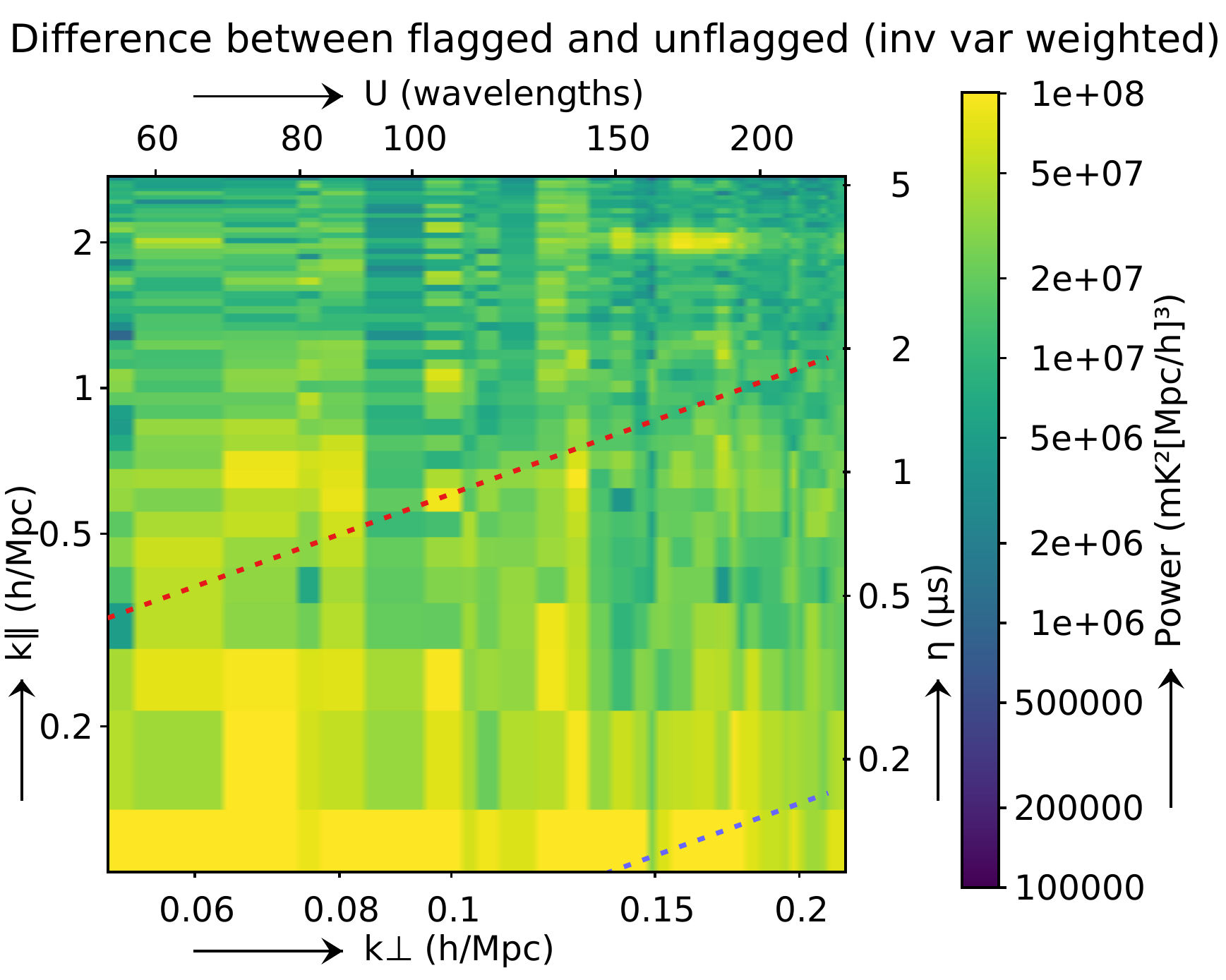}%
\includegraphics[width=8cm]{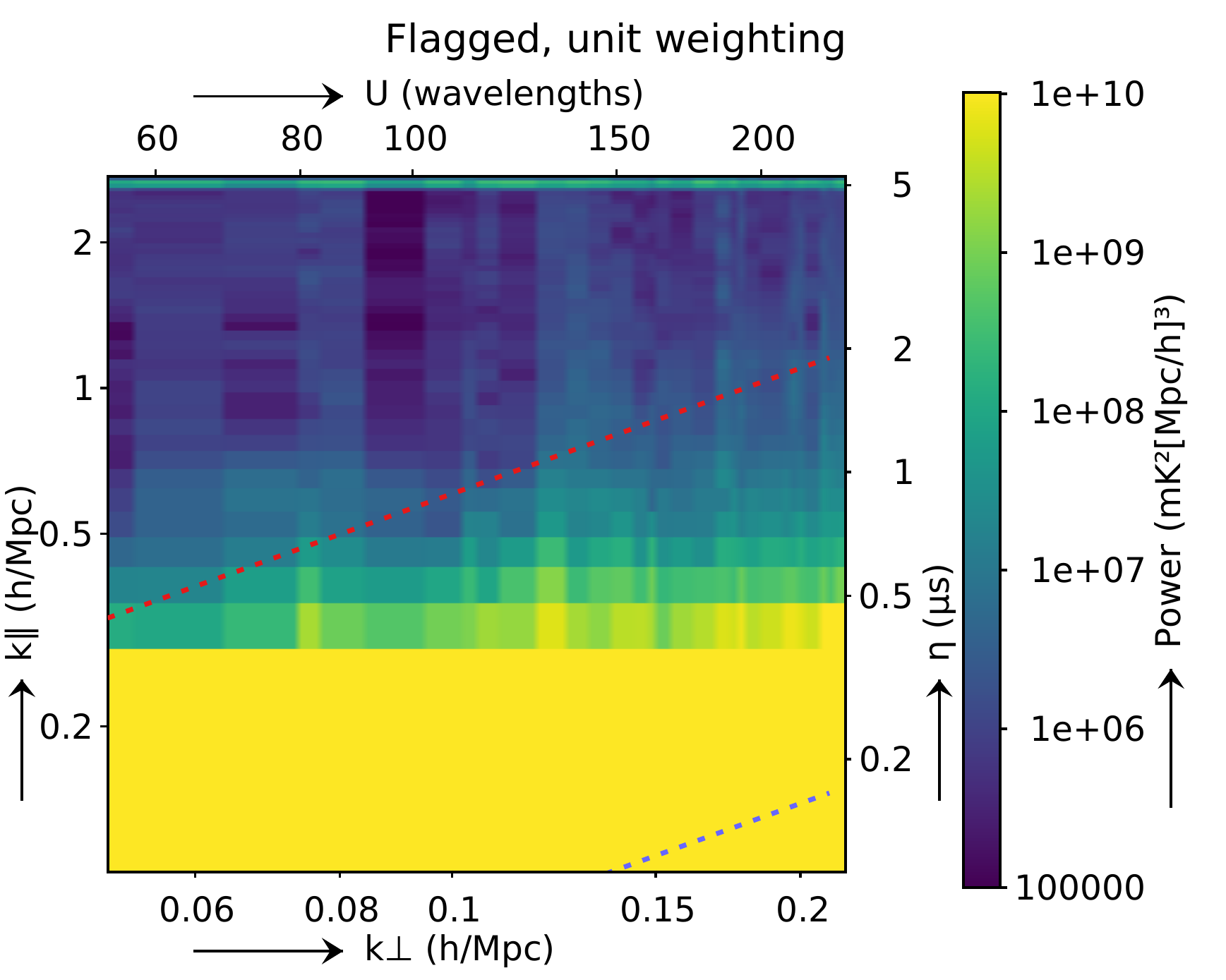}%
\caption{Cylindrically-averaged power spectra from simulated noise-less data. Top-left: from simulated data without any flags; Top-right: from simulated data after applying the RFI flags from a real observation and using inverse-variance weighting; Bottom-left: difference between non-flagged (top-left) and flagged data (top-right); bottom-right: same as top-right, but giving visibilities equal weight, independent of how many visibilities were flagged before averaging. The red dashed line indicates the $k$-modes that correspond to the horizon; the blue dashed line corresponds to a 5\degree field of view.}
	\label{fig:spherical-ps}
\end{figure*}

The spectral fluctuations caused by data flagging causes some part of the power of the foregrounds to have spectral fluctuations that correspond with the redshifted signals from the Epoch of Reionization. For example, the cylindrically-averaged power spectra corresponding to the simulated clean and flagged data in Fig.~\ref{fig:spherical-ps} show that flagging causes a significant increase of power at high $k_\parallel$, above the foreground wedge. In the next sections, we explain the source of the fluctuations and describe methods to mitigate them.

\subsection{Temporal and spectral averaging} \label{sec:averaging}
Data averaging is a common step in radio astronomy: it is implemented in several pipelines, including the LOFAR EoR pipeline, and can be performed with several tools (e.g. \textsc{casa}, \textsc{dppp} and \textsc{cotter}). In this study we find that the standard method of averaging data increase the spectral fluctuations caused by flagging.

The way averaging in time and frequency are implemented in the aforementioned tools is by binning samples that are close together in time and frequency. When samples are flagged, they are excluded from the bin, causing the time-frequency bin to not exactly be equal to the mean of the visibility function of the sky. It is effectively equal to interpolating the flagged sample and replacing it by the average of the mean of the other samples in that bin. While the error made by this method (the difference between the interpolated value and the true value of the interpolated visibility) is small and mostly negligible in regular radio-astronomical data analysis, cosmological 21-cm power spectrum analyses are extremely sensitive to spectral fluctuations, and can be negatively affected by this effect.

To form images from visibilities, samples are gridded on a regular $uv$ plane. Similar to averaging, data is also binned during data gridding, and samples that are missing due to RFI will effect the imaging in a similar way. Flagged samples can therefore cause spectral fluctuations both during data averaging as well as during gridding.

We also find that another technical detail is relevant to the excess flagging power. In standard radio interferometric data, each visibility has an associated weight stored with it. This weight is taken into account during calibration and imaging, which results in inverse-variance weighted, least-squares calibration solutions and inverse-variance weighted images, and therefore in an inverse-variance weighted power spectrum. When averaging data, the visibility weights are normally updated: when half the input samples in an averaging bin are flagged, the averaged visibility will get half the weight of a fully averaged sample. Unfortunately, weighting samples based on the RFI flags during imaging will cause differences between the $uv$-coverage at different frequencies. Similar to flagging and averaging, having different $uv$-coverages at different frequencies can lead to spectral fluctuations. The effect of non-smooth weights over frequency will, when visibilities are averaged or binned onto a $uv$-grid, lead to effects that are similar to the effect of missing samples. On first order, it changes the centroid of a $uv$-cell with unevenly distributed weights, causing fluctuations over frequency. 
We will compare the inverse-variance weighting method to unitary weighting of the visibilities, in which all visibilities are given equal weight independently of how many samples were flagged before averaging.

\subsection{Improved interpolation during averaging and gridding} \label{sec:interpolation}
As discussed in a previous section, during standard radio-interferometric data averaging, flagged samples are interpolated and replaced by the mean of the time-frequency bin they are in. This binning interpolation method contributes to the flagging excess power, because a bin with flagged samples can have a biased average that results in spectral fluctuations. In this paper we test an improved interpolation scheme.

For our improved interpolation scheme, before averaging we replace flagged samples by the windowed, Gaussian-weighted average of unflagged samples:
\begin{equation} \label{eq:interpolation}
 V'(i,j)=\frac{
	\sum_{k,l} W(k,l) F(i+k,j+l) V(i+k,j+l)
}{
	\sum_{k,l} W(k,l) F(i+k,j+l)
},
\end{equation}
with $W(k,l)=\exp(-0.5(k^2+l^2)/\sigma^2)$, the two-dimensional Gaussian function with width parameter $\sigma$; $V(i,j)$ the complex visibility for timestep $i$ and channel $j$; and $F(i,j)$ the flag status for sample $(i,j)$: $0$ if it is flagged and $1$ otherwise. The sums in Eq.~\eqref{eq:interpolation} are over $l,k \in [-\frac{1}{2}(N-1) : \frac{1}{2}(N-1)]$, with $N$ the (odd) size of the window. We have chosen $\sigma$ to be the width of one timestep/channel (corresponding to a temporal $\sigma$ of 2s and a spectral $\sigma$ of 3 kHz), which we find is large enough to calculate a representable visibility for missing samples, and at the same time small enough to avoid the need of a computationally-expensive large window.

\subsection{Forward modelling}
As long as the high-resolution RFI detection flags are stored along with the averaged data, it is possible to forward predict the sky model, including the effect of data flagging and averaging. This requires storing the high-resolution flags, predicting the model at high resolution and propagating the high resolution model data through the same flagging and averaging steps. This would cancel the excess flagging power associated with the modelled sky sources. However, doing so requires prediction of the sky model at the time and frequency resolution at which flagging is performed, which is computationally expensive. For LOFAR, predicting the foreground model at low resolution costs already approximately half of our total computational budget. Predicting at high resolution (> 100 times more visibilities) is therefore expensive with current techniques, although optimizations might be possible. Prediction at low resolution is however already feasible. A low-resolution prediction will not fully take into account the loss of flagged high-resolution samples, but will remove most of the foreground power before gridding, and this will therefore reduce the excess power associated with the gridding step. We will assess to what level low-resolution modelling mitigates the flagging excess power. 

\section{Results} \label{sec:results}

\begin{figure}
\includegraphics[width=\columnwidth]{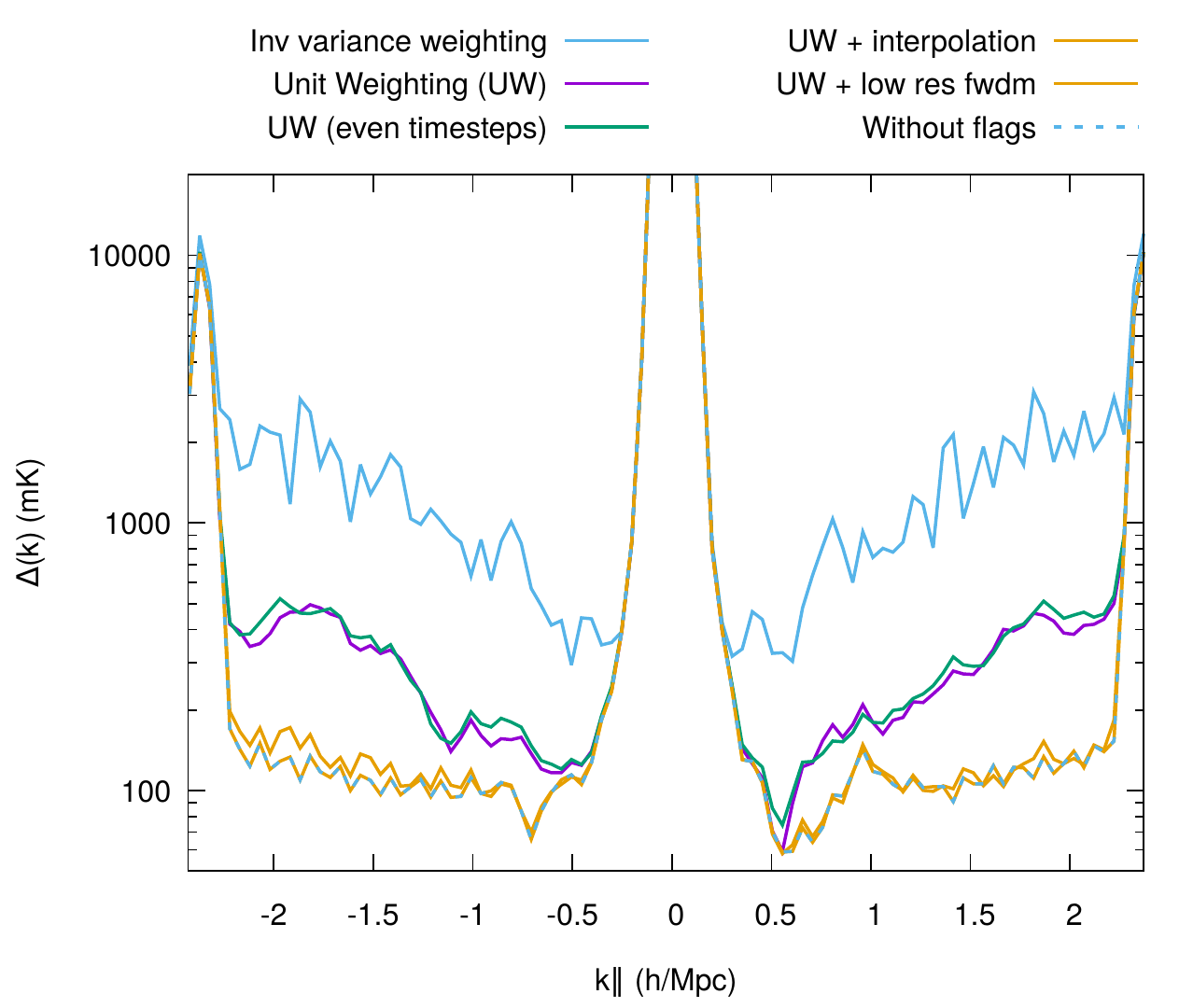}%
	\caption{Power at $k_\perp=0.1$ $h$/Mpc without applying GPR, showing power as function of $k_\parallel$ with and without flagging. Solid lines are from simulated data that includes flagged RFI, the dashed line is without any flagged samples. }
	\label{fig:kpar-spectra-pstransform}
\end{figure}

\begin{figure}
\includegraphics[width=\columnwidth]{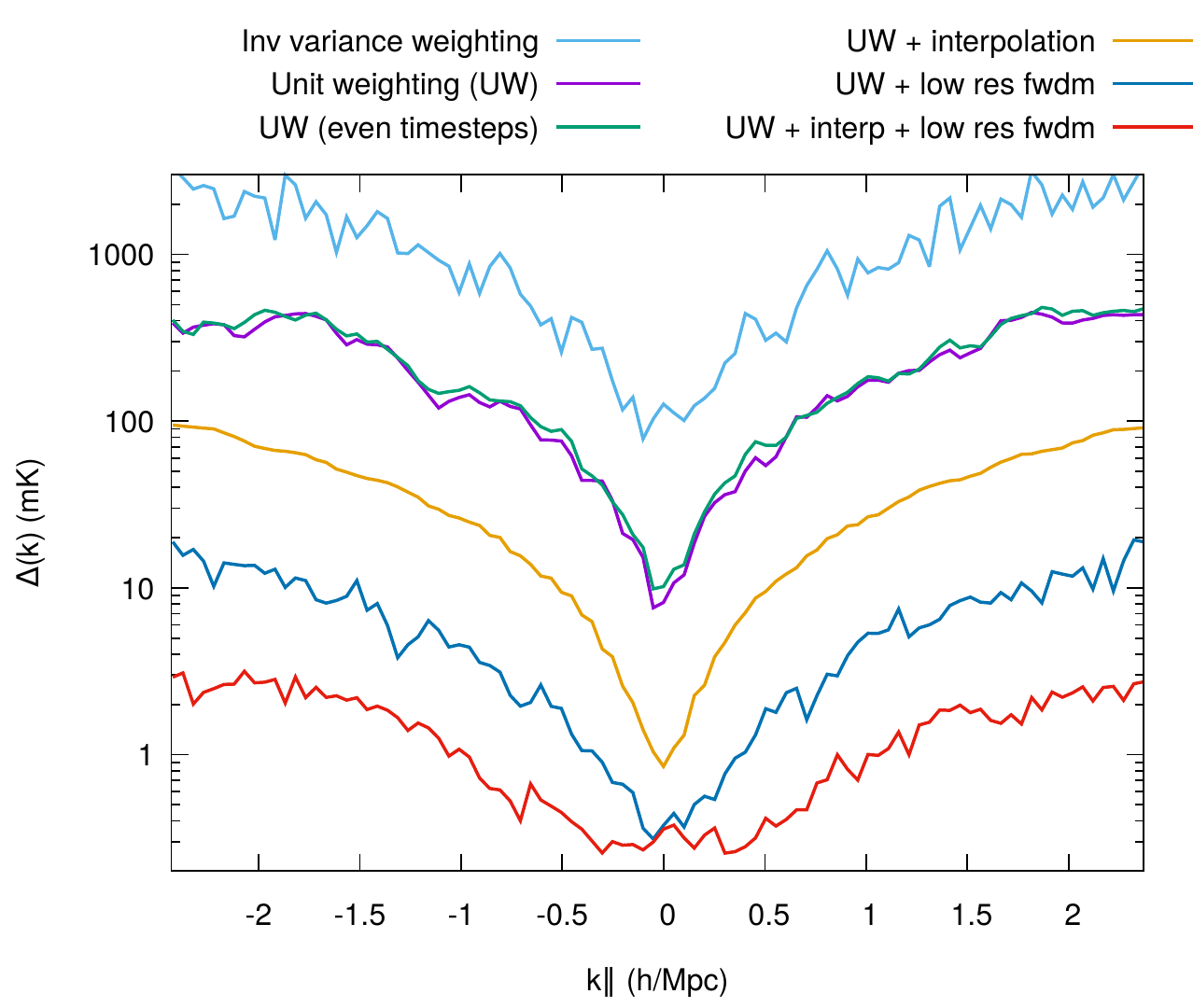}%
	\caption{Difference power at $k_\perp=0.1$ $h$/Mpc showing the excess power introduced by data flagging for the same cases as in Fig.~\ref{fig:kpar-spectra-pstransform}. }
	\label{fig:kpar-difference-pstransform}
\end{figure}

Fig.~\ref{fig:spherical-ps} shows cylindrically-averaged power spectra from the simulated data. Because no foreground removal strategy has been applied yet, the power spectra are dominated by power from the spectrally-smooth foregrounds. The power in the plot decreases rapidly with increasing $k_\parallel$. There is a horizontal line of high power visible at $k_\parallel \approx2.5$ $h$/Mpc, which is caused by missing channels at the 200~kHz LOFAR subband edges. The poly-phase filter that forms the subbands aliases the signals from other channels into these channels, and these channels have to be removed in observations. We have therefore also removed these edge channels in the simulations.

The top-right panel in Fig.~\ref{fig:spherical-ps} shows the cylindrically-averaged power spectra from data that includes RFI flags. It can be seen that the spectral fluctuations caused by flagging result in excess power at high $k_\parallel$-values. As can be seen in the bottom-left image of Fig.~\ref{fig:spherical-ps}, which shows the power spectrum of the data cube difference with and without flagging, the flagging excess power affects all $k_\parallel$-values with approximately equal power. It therefore also contaminates the power spectrum window in which the epoch of reionization signals are most easily detectable. 

Certain effects can be absorbed in calibration solutions. To validate whether calibration affects the excess power, we have also performed calibration of the simulated data. This results in fact in a slightly increased excess power: the excess power is not suppressed by calibration.

\paragraph*{Visibility weighting results:}
So far, the visibilities were inverse-variance weighted, i.e., the number of unflagged visibilities before averaging is propagated into the weight of an individual averaged visibility. The bottom-right panel of Fig.~\ref{fig:spherical-ps} shows the power spectrum from the same data after giving all visibilities the same weight. The excess flagging power has decreased quite significantly compared to the inverse-variance visibility weighted plot, but excess power is still visible. Similar to the inverse-variance weighting case, the power spectrum from unitary-weighted visibilities is affected with approximately similar power at all modes. The reason that inverse variance increases flagging excess power, is that the inverse-variance weights are dependent on the number of flagged samples, and therefore cause the $uv$-coverage to be different for different frequencies. On first order, having spectrally fluctuating weights causes the centroid of $uv$-cells to be different for different frequencies and therefore results in small fluctuations in the gridded visibilities.

Fig.~\ref{fig:kpar-spectra-pstransform} shows $k_\perp=0.1$ $h$/Mpc slices through the cylindrically-averaged power spectra of Fig.~\ref{fig:spherical-ps}, converted to dimensionless power $\Delta(k)=\sqrt{P k^3 /(2\pi^2)}$, with $k=\sqrt{k^2_\parallel + k^2_\perp}$. Similarly, Fig.~\ref{fig:kpar-difference-pstransform} shows the difference between the non-flagged ``ground truth'' data and various cases that include flagging. The difference plot is constructed by subtracting the affected visibilities from the ground-truth visibilities in gridded uv-space before calculating the power spectrum. The values in these plots can be compared to the expected value of the power in the 21-cm signal fluctuations, which is typically predicted to be several to tens of mK at these redshifts and $k$-values (e.g. \citealt{greig-2015-21cmmc}). Because of the conversion to dimensionless power, the power (and excess flagging power) increases toward higher $|k_\parallel|$ values. The power levels are still relatively high because no foregrounds were yet subtracted. With inverse variance weighting, the excess flagging amplitude at $k_\perp=0.1$ $h$/Mpc ranges from approximately 100 to 3000\,mK at low and high $k_\parallel$, respectively. This decreases to 10 to 400\,mK when using unitary visibility weighting.

Another correction we try is to grid the visibilities on their true centroid during the gridding process. The LOFAR DPPP software stores metadata in the measurement set from which the true visibility centroid position can be inferred. However, we find that such a correction results in no improvement.

\begin{figure}
\includegraphics[width=\columnwidth]{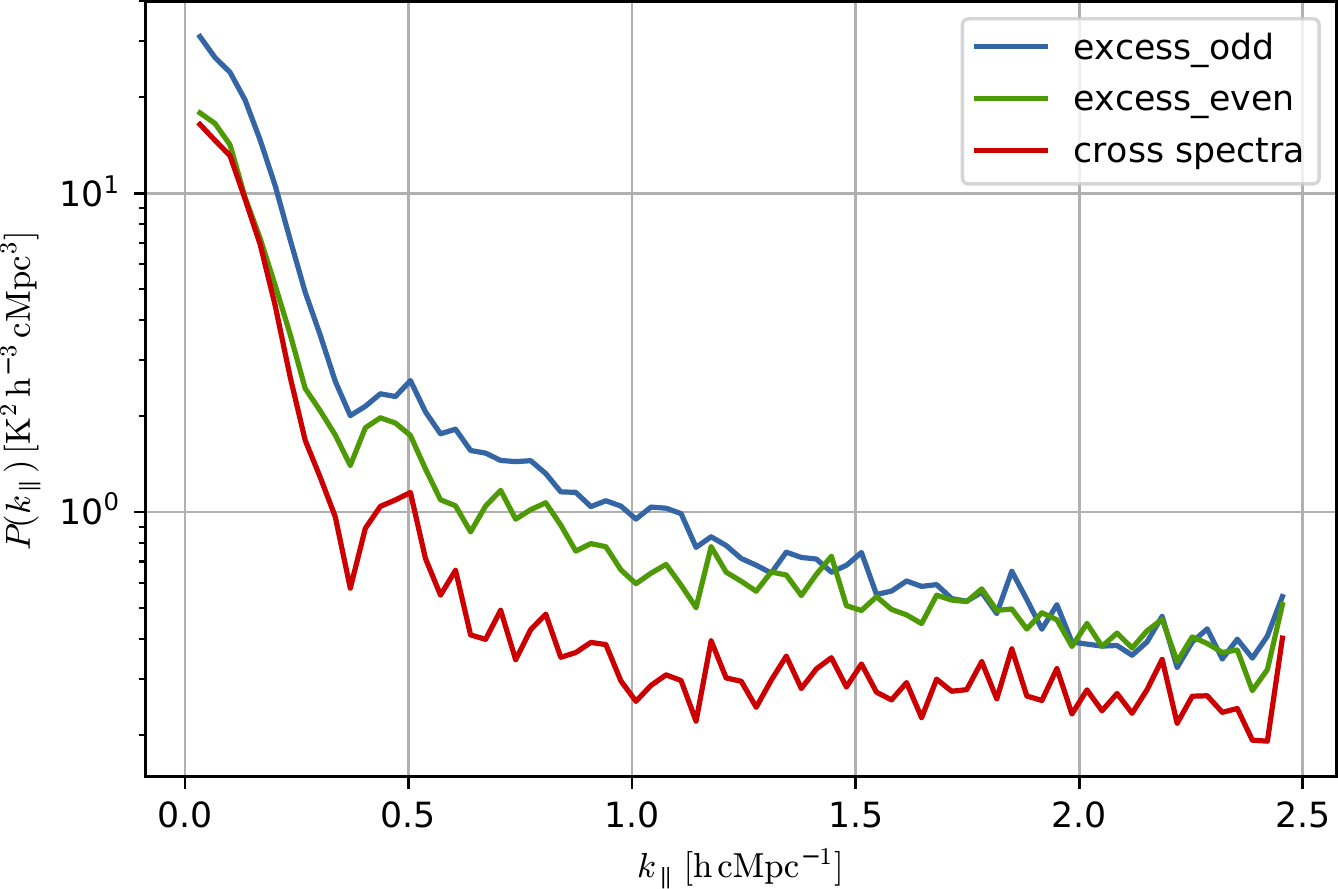}%
	\caption{2D power-spectra averaged over all $k_x$ and $k_y$, comparing the odd and even timestep data, as well as its cross spectrum. At small $k_\parallel$-values, the odd and even sets show correlated excess power. At $k_\parallel \gtrsim 0.3$ $h$/Mpc, the cross spectra is about a factor of 2 below the individual spectra, implying that the flagging artefacts correlate partially between the odd and even sets. }%
	\label{fig:odd-even-excess}
\end{figure}

\begin{figure}
\includegraphics[width=\columnwidth]{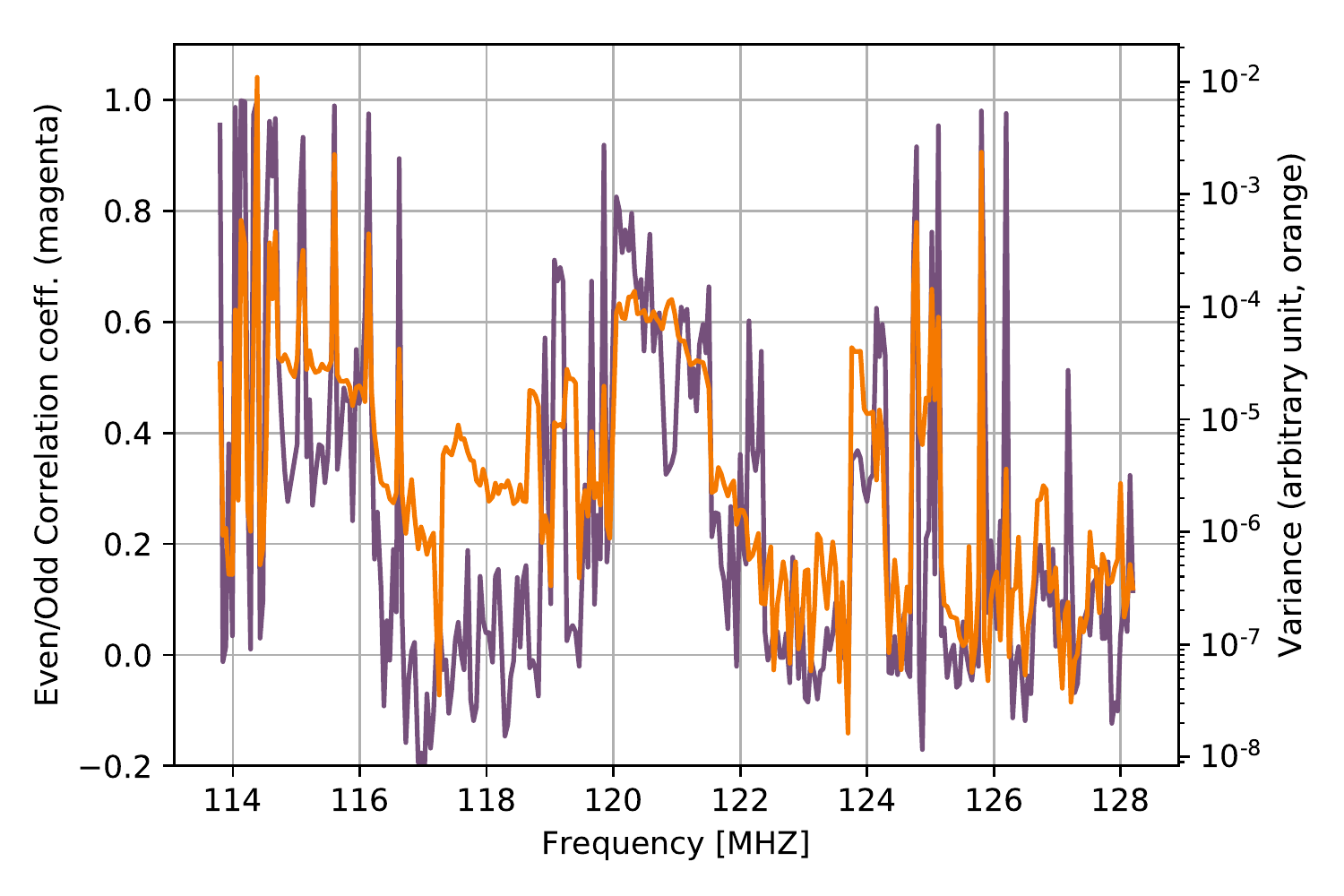}%
	\caption{Correlation between the odd and even cases (magenta line) drawn together with the variance (orange line). Bands with higher RFI excess power, and therefore higher RFI occupancy, show a stronger correlation between odd and even. This implies that stronger sources are more consistently present.}%
	\label{fig:odd-even-frequency-correlation}
\end{figure}

\paragraph*{Gaussian interpolation results:}
In \S\ref{sec:interpolation} we have described an interpolation method to replace flagged samples by a Gaussian-weighted average of unflagged samples. This type of interpolation was implemented in the \textsc{dppp} software.
To test this method, we construct a simulated high-resolution LOFAR data cube including flags as before. Subsequently, we interpolate flagged values with a window size $N=15$ and a Gaussian kernel size of $\sigma=1$. If for a particular sample all samples within the corresponding window are flagged, the sample is not interpolated and the corresponding output sample is flagged. 

Even though the interpolation step is performed at high resolution, it is computationally insignificant compared to the RFI detection step and reading of the data. 

\begin{figure*}
\includegraphics[width=17.5cm]{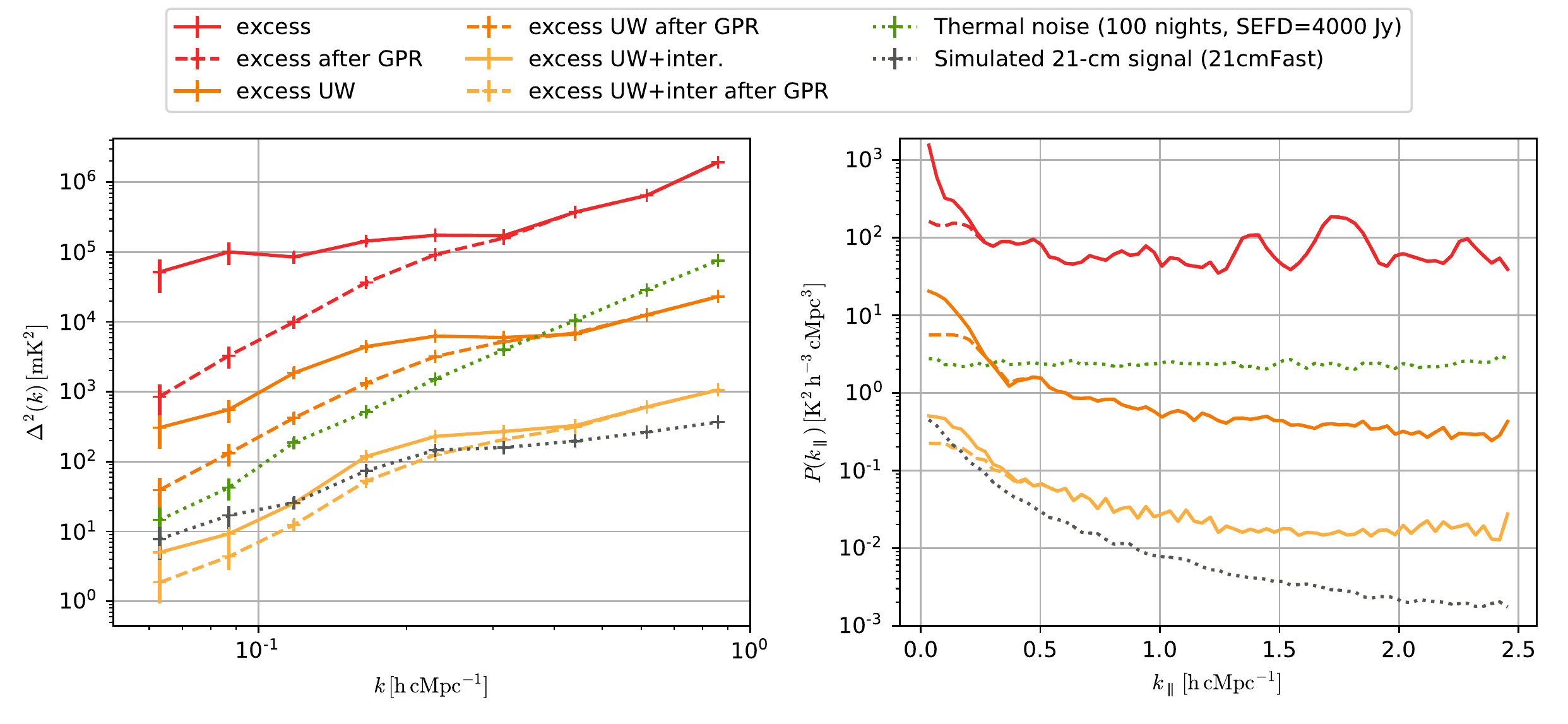}%
	\caption{Impact on the excess power of the GPR procedure. Left: Spherically averaged
power-spectra. Right: 2D power-spectra averaged over all $k_x$ and $k_y$. The excess power is shown for three different processing
setups, before (solid line) and after applying GPR (dashed line). For
comparison, we also show the thermal noise power spectra corresponding to about
100 nights of LOFAR-HBA observation (assuming an SEFD=4000 Jy), and the power
spectra of a simulated 21-cm signal.}
	\label{fig:gpr}
\end{figure*}

The power spectrum results for this interpolation scheme are visualized in Fig.~\ref{fig:kpar-spectra-pstransform} and Fig.~\ref{fig:kpar-difference-pstransform}. Compared to normal averaging with unit weighted visibilities, Gaussian interpolation reduces the excess flagging power by over an order of magnitude. At small $k_\parallel$ values, the excess noise decreases to a level of approximately one mK. With this decrease in excess power, together with further decomposition of the data by GPR and the decrease caused by averaging multiple nights together, the flagging excess power will no longer prevent detection of 21-cm signals from the EoR. 

\paragraph*{Forward modelling results:}
Fig.~\ref{fig:kpar-difference-pstransform} includes the result of subtracting the predicted sky model from the low resolution data. After low-resolution prediction, the residual excess power varies from 0.3 to 19 mK. The residual excess power is caused by the fact that a low-resolution prediction does not match the high-resolution predicted data precisely, because of the combination of flagging and averaging. Compared to high-resolution interpolation, the low-resolution sky model subtraction reduces the excess power by approximately a factor of 2 to 5 at low and high $k_\parallel$, respectively. Considering that a low-resolution predict removes power before the gridding operation, and can not model the excess fluctuations that are arising from averaging high-resolution data, this result implies that most of the flagging excess power arises during gridding of the low-resolution data on the $uv$-plane, and not during the first stage of high-resolution data averaging. Although the low-resolution data has considerably fewer flags because of data averaging, in which flagged values are replaced by the bin average, a number of samples remains flagged after averaging. Those flags arise when an averaging bin (or interpolation kernel) was completely flagged in the high resolution data. It is those flags that cause most of the excess power.

This result also implies that flagging at lower resolution to skip the time-frequency averaging, will not avoid the excess noise: although the averaged visibilities do no longer have biases due to the high-resolution flagging, the gridding of visibilities in bins would still be affected by missing low-resolution samples. Moreover, flagging at a lower resolution is less effective, causing more false negative.

Low-resolution prediction cannot solve inaccuracies that were petrified into the data during high-resolution processing. However, by using Gaussian interpolation of the high-resolution data instead of data averaging, it is possible to reduce the excess power further. As shown in Fig.~\ref{fig:kpar-difference-pstransform}, the combination of uniform weighting, high-resolution interpolation and low-resolution forward modelling results in the lowest excess power, being $< 1\,$mK at $k_\parallel < 1$\,h/Mpc.

\subsection{Systematic nature of excess flagging power}
So far, we have simulated a single night of observation. The single-night results indicate that, if not mitigated, the excess flagging power can be above likely predictions of the 21-cm signal power. However, if this power would decrease with time similar to the system noise, it will be an order of magnitude lower for a 100-night data analysis. When using unitary visibility weighting with no further mitigation, the results show that the excess power will in that case (mostly) not prevent detection of the 21-cm signals. We investigate therefore whether excess flagging power indeed behaves noise-like or has a systematic nature.

Certain transmitting sources of interference will occupy the same frequencies consistently, while other sources of interference might behave more erratic, occupying random timeslots or channels. Examples of sources for such interference are lightning, solar flares and sparking devices. To assess whether the excess flagging power is systematically present or whether it averages down with time, we split our data in even and odd timesteps. This approach is taken instead of constructing a second night of simulated data, because simulating a second high-resolution observation with real RFI flags from a second night is practically difficult, mainly because storing (another) large volume of data on our EoR cluster would interfere with the running LOFAR EoR observations. Another approach would be to analyse the time-dependence by splitting the observation halfway in time. This would be slightly more representative, because the even and odd timesteps are more likely to correlate than the first and second half of the observation. However, doing so resulted in excess noise that is difficult to interpret, because the $uv$-coverage of the first and second halves are different.

If the excess power is not systematic, using half the data should increase the power (in units of mK) by approximately a factor of $\sqrt{2}$, comparable with the system noise. However, as shown in Figs.~\ref{fig:kpar-spectra-pstransform} and \ref{fig:kpar-difference-pstransform}, the data do not agree with this hypothesis: the excess power in the even-timestep set is at the same level as in the full data. This implies that this source of excess power is systematic and could therefore be problematic for EoR experiments if not mitigated.

To further analyse the correlation of the flagging excess power over time, we construct a cross power spectrum between odd and even timestep sets. If the excess power is uncorrelated in time, a cross spectrum between the two sets should decrease the excess power substantially. In Fig.~\ref{fig:odd-even-excess} it is shown that for $k\gtrsim 0.5$ $h$/Mpc, the cross-spectrum is approximately a factor of two below the individual spectra. Fig.~\ref{fig:odd-even-frequency-correlation} shows the correlation over frequency, which indicates that an increase in excess power at a particular frequency also causes an increased correlation. A possible explanation for this is that stronger sources of interference are consistently present at the same frequency, while weaker, transient sources of interference and false positive flags do not transmit/occur at one specific frequency.

\subsection{Impact of GPR on the excess power}
By applying GPR on simulated data sets, one can analyse what would be the impact
of running GPR on RFI affected observation, and if part of the flagging excess
power can be modelled by GPR as part of the foregrounds mode-mixing. In Fig.~\ref{fig:gpr},
we analyse excess image cubes that are constructed by subtracting an unflagged
from a flagged data cube. As before, noiseless simulations are used.
GPR is able to model part of the excess power at lower
k-values. This is particularity the case when using inverse-variance weighting
and without interpolation, for which a reduction in excess power of more than
one order of magnitude at small $k$ is observed. When using unit-weighting and
with interpolation, the frequency-coherence of the excess power is reduced, and
as is the impact of the GPR. Nevertheless at this stage the excess power is
considerably reduced and is at the level of the 21-cm signal. Flagging the
subbands which are most affected by RFI (see e.g. Fig.~\ref{fig:odd-even-frequency-correlation})
and subtracting the low-resolution forward model before gridding could reduce it even more.

\section{Conclusions} \label{sec:conclusions}
For the LOFAR EoR case we have shown that RFI flagging followed by data averaging and gridding, with no further mitigation, causes excess power that is significantly above the expected 21-cm signals and does not considerably average down over time. In order to achieve a detection of 21-cm signals, the excess power can be significantly decreased by: i) using the same weight for all visibilities, instead of propagating the number of visibilities in the averaging bin (i.e. inverse-variance weighting); ii) forward modelling (``prediction'') of the data at low resolution; and iii) Gaussian interpolation of missing samples prior to any data averaging, instead of the commonly-used method of replacing a sample by the mean of the averaging bin. Furthermore, a part of the excess power behaves statistically as normally-distributed foregrounds, and Gaussian process regression can remove about another factor of two of the excess power. Subbands that see a larger contamination of RFI show stronger excess flagging power, and a final mitigation strategy is to select bands that are less contaminated by RFI. Together these techniques reduce the excess flagging power by approximately three orders of magnitude. In particular, inside the EoR window at $k=0.1$\,$h$/Mpc the power is reduced from about 200 to 0.3\,mK, where the 21-cm signals are expected to be on the order of several mK.

The presented techniques have been implemented in the pipeline of the LOFAR EoR project. Subtraction of the low-resolution forward model is part of the direction-dependent calibration scheme. This scheme uses Sagecal CO \citep{yatawatta-co-2016} to subtract the model from the averaged data including direction-dependent effects.

Because the excess power scales linearly with the foreground power, mitigation of the flagging excess power is even more critical for fields with higher sky temperature. One of the fields that is also targetted as part of the LOFAR EoR project, is a field that is centred on 3C 196, a bright ($\sim80$\,Jy) quasar at corresponding frequencies \citep{scaife-heald-2012}, and the temperature of this field of view is therefore about an order of magnitude above more quiet fields.

Although this issue was analysed in the context of the LOFAR EoR project, it can be assumed that our conclusions are equally applicable to other EoR experiments. Flagging excess power scales linearly with the number of flagged samples, and although telescopes such as the Murchison Widefield Array are in a more benign RFI environment compared to LOFAR, the percentage of flagged samples due to RFI are comparable: in \citet{offringa-2015-mwa-rfi}, the overall percentage of RFI observed with the MWA is 1.1\%. For the LOFAR bandwidth used in this work this value is 1.12\% (see Table~\ref{tbl:observation}). One of the reasons for this relatively low occupation of RFI compared to the MWA is the high time and frequency data recording resolution of LOFAR \citep{offringa-lofar-environment-2013}. Delay delay-rate filtering \citep{parsons-2009} to remove foregrounds as performed by the PAPER EoR team is similar to interpolation and/or forward modelling, and is likely to decrease the flagging excess power as well, depending on the resolution of the data at which it is applied. 

These conclusions are also important for the SKA telescope, for which the data rate is so high that early averaging of data is mandatory, and the full resolution data at which the RFI is detected will not be kept. For a future SKA EoR experiment, it is probably necessary to implement a scheme similar to the Gaussian interpolation introduced in this work. It is also important to store the high-resolution flags when averaging the data, so that any residual flagging excess power from modelled sources can be forward modelled as accurately as is computationally allowed.

In some cases calibration can remove unmodelled effects, such as cable reflections or beam changes. In the case of flagging excess power, we found that the observed excess power is not absorbed in calibration solutions. This is expected, because RFI flagging is inherently a baseline-based effect, and will be variable on time and frequency scales smaller than the solution interval. This is particularly the case when the solutions are constrained to be spectrally smooth, as is desirable to avoid suppression of the unmodelled 21-cm signals \citep{patil-2016} and to avoid calibration-induced excess power \citep{barry-2016}.

We have shown how interpolation using a Gaussian kernel before time and frequency averaging can help in reducing excess power. It is worth mentioning that this interpolation technique does not only benefit EoR studies, but will in general result in more accurately averaged visibilities compared to the de facto method of replacing missing samples by the mean of the averaging bin. This is because for a particular sample, the Gaussian interpolated visibility will generally represent the interpolated visibility more closely than the mean of a time-frequency averaged bin. With the de facto averaging method, an averaging bin is generally not centred on the interpolated visibility, and visibilities are weighted equally independent of their Euclidean time-frequency distance to the interpolated visibility. Because of the small magnitude of the effect that averaging of flagged samples has, improved interpolation will likely be of inconsiderable small magnitude for most science cases. Nevertheless, it might still be relevant for reaching high dynamic ranges, for example in continuum imaging. Because of the small computational cost of the method, it is straightforward to implement.

\section*{Acknowledgements}
A.\,R.\,Offringa acknowledges financial support from the European Research Council under ERC Advanced Grant LOFARCORE 339743. FGM and LVEK would like to acknowledge support from a
SKA-NL Roadmap grant from the Dutch ministry of OCW.
We thank M. Mevius for helpful discussions.


\DeclareRobustCommand{\TUSSEN}[3]{#3}

\bibliographystyle{mnras}
\bibliography{references}

\begin{thebibliography}{}
\makeatletter
\relax
\def\mn@urlcharsother{\let\do\@makeother \do\$\do\&\do\#\do\^\do\_\do\%\do\~}
\def\mn@doi{\begingroup\mn@urlcharsother \@ifnextchar [ {\mn@doi@}
  {\mn@doi@[]}}
\def\mn@doi@[#1]#2{\def\@tempa{#1}\ifx\@tempa\@empty \href
  {http://dx.doi.org/#2} {doi:#2}\else \href {http://dx.doi.org/#2} {#1}\fi
  \endgroup}
\def\mn@eprint#1#2{\mn@eprint@#1:#2::\@nil}
\def\mn@eprint@arXiv#1{\href {http://arxiv.org/abs/#1} {{\tt arXiv:#1}}}
\def\mn@eprint@dblp#1{\href {http://dblp.uni-trier.de/rec/bibtex/#1.xml}
  {dblp:#1}}
\def\mn@eprint@#1:#2:#3:#4\@nil{\def\@tempa {#1}\def\@tempb {#2}\def\@tempc
  {#3}\ifx \@tempc \@empty \let \@tempc \@tempb \let \@tempb \@tempa \fi \ifx
  \@tempb \@empty \def\@tempb {arXiv}\fi \@ifundefined
  {mn@eprint@\@tempb}{\@tempb:\@tempc}{\expandafter \expandafter \csname
  mn@eprint@\@tempb\endcsname \expandafter{\@tempc}}}

\bibitem[\protect\citeauthoryear{Barry, Hazelton, Sullivan, Morales  \&
  Pober}{Barry et~al.}{2016}]{barry-2016}
Barry N.,  Hazelton B.,  Sullivan I.,  Morales M.~F.,   Pober J.~C.,  2016,
  \mn@doi [MNRAS] {10.1093/mnras/stw1380}, 461, 3135

\bibitem[\protect\citeauthoryear{Beardsley et~al.,}{Beardsley
  et~al.}{2016}]{beardsley-2016}
Beardsley A.~P.,  et~al., 2016, ApJ, 833, 102

\bibitem[\protect\citeauthoryear{Bernardi, {\TUSSEN{Bruyn}{De}{de}}~Bruyn,
  Harker  et~al.}{Bernardi et~al.}{2010}]{bernardi-wsrt-foregrounds-II-2010}
Bernardi G.,  {\TUSSEN{Bruyn}{De}{de}}~Bruyn A.~G.,  Harker G.,   et~al., 2010,
  A\&A, 522, A67

\bibitem[\protect\citeauthoryear{Callingham et~al.,}{Callingham
  et~al.}{2017}]{callingham-2017}
Callingham J.~R.,  et~al., 2017, ApJ, 836, 174

\bibitem[\protect\citeauthoryear{Carroll et~al.,}{Carroll
  et~al.}{2016}]{carroll-2016}
Carroll P.~A.,  et~al., 2016, \mn@doi [MNRAS] {10.1093/mnras/stw1599}, 461,
  4151

\bibitem[\protect\citeauthoryear{Chapman et~al.,}{Chapman
  et~al.}{2013}]{chapman-2013}
Chapman E.,  et~al., 2013, \mn@doi [MNRAS] {10.1093/mnras/sts333}, 429, 165

\bibitem[\protect\citeauthoryear{Datta, Bowman  \& Carilli}{Datta
  et~al.}{2010}]{datta-2010-eor-foreground-subtraction}
Datta A.,  Bowman J.~D.,   Carilli C.~L.,  2010, ApJ, 724, 526

\bibitem[\protect\citeauthoryear{{Ewall-Wice}, {Dillon}, {Liu}  \&
  {Hewitt}}{{Ewall-Wice} et~al.}{2017}]{ewall-wice-2017}
{Ewall-Wice} A.,  {Dillon} J.~S.,  {Liu} A.,   {Hewitt} J.,  2017, \mn@doi
  [\mnras] {10.1093/mnras/stx1221}, \href
  {http://adsabs.harvard.edu/abs/2017MNRAS.470.1849E} {470, 1849}

\bibitem[\protect\citeauthoryear{Franzen, Jackson, Offringa  et~al.}{Franzen
  et~al.}{2016}]{franzen-2016}
Franzen T. M.~O.,  Jackson C.~A.,  Offringa A.~R.,   et~al., 2016, \mn@doi
  [MNRAS] {10.1093/mnras/stw823}, 459, 3314

\bibitem[\protect\citeauthoryear{Furlanetto, Oh  \& Briggs}{Furlanetto
  et~al.}{2006}]{furlanetto-2006}
Furlanetto S.~R.,  Oh S.~P.,   Briggs F.~H.,  2006, \mn@doi [Physics Reports]
  {https://doi.org/10.1016/j.physrep.2006.08.002}, 433, 181

\bibitem[\protect\citeauthoryear{Ghosh, Mertens  \& Koopmans}{Ghosh
  et~al.}{2018}]{ghosh-2018}
Ghosh A.,  Mertens F.~G.,   Koopmans L. V.~E.,  2018, \mn@doi [MNRAS]
  {10.1093/mnras/stx2959}, 474, 4552

\bibitem[\protect\citeauthoryear{Greig \& Mesinger}{Greig \&
  Mesinger}{2015}]{greig-2015-21cmmc}
Greig B.,  Mesinger A.,  2015, \mn@doi [MNRAS] {10.1093/mnras/stv571}, 449,
  4246

\bibitem[\protect\citeauthoryear{{\TUSSEN{Haarlem}{Van}{van}}~Haarlem
  et~al.,}{{\TUSSEN{Haarlem}{Van}{van}}~Haarlem et~al.}{2013}]{lofar-2013}
{\TUSSEN{Haarlem}{Van}{van}}~Haarlem M.~P.,  et~al., 2013, \mn@doi [A\&A]
  {10.1051/0004-6361/201220873}, 556, A2

\bibitem[\protect\citeauthoryear{Harker et~al.,}{Harker
  et~al.}{2009}]{harker-2009}
Harker G.,  et~al., 2009, \mn@doi [MNRAS] {10.1111/j.1365-2966.2009.15081.x},
  397, 1138

\bibitem[\protect\citeauthoryear{Hurley-Walker et~al.,}{Hurley-Walker
  et~al.}{2017}]{hurley-walker-2017-gleam}
Hurley-Walker N.,  et~al., 2017, \mn@doi [MNRAS] {10.1093/mnras/stw2337}, 464,
  1146

\bibitem[\protect\citeauthoryear{{Iliev}, {Shapiro}, {Ferrara}  \&
  {Martel}}{{Iliev} et~al.}{2002}]{liev-2002}
{Iliev} I.~T.,  {Shapiro} P.~R.,  {Ferrara} A.,   {Martel} H.,  2002, \mn@doi
  [\apjl] {10.1086/341869}, \href
  {http://adsabs.harvard.edu/abs/2002ApJ...572L.123I} {572, L123}

\bibitem[\protect\citeauthoryear{Jacobs et~al.,}{Jacobs
  et~al.}{2016}]{jacobs-2016}
Jacobs D.~C.,  et~al., 2016, ApJ, 825, 114

\bibitem[\protect\citeauthoryear{Jeli\'c et~al.,}{Jeli\'c
  et~al.}{2008}]{jelic-lofar-foregrounds-2008}
Jeli\'c V.,  et~al., 2008, \mn@doi [MNRAS] {10.1111/j.1365-2966.2008.13634.x},
  389, 1319

\bibitem[\protect\citeauthoryear{Kazemi, Yatawatta, Zaroubi, Labropoulos,
  \TUSSEN{Bruyn}{De}{de}, Koopmans  \& Noordam}{Kazemi
  et~al.}{2011}]{kazemi-2011-sagecal}
Kazemi S.,  Yatawatta S.,  Zaroubi S.,  Labropoulos P.,  \TUSSEN{Bruyn}{De}{de}
  A.~G.,  Koopmans L. V.~E.,   Noordam J.,  2011, MNRAS, 414, 1656

\bibitem[\protect\citeauthoryear{Liu, Tegmark  \& Zaldarriaga}{Liu
  et~al.}{2009}]{liu-2009}
Liu A.,  Tegmark M.,   Zaldarriaga M.,  2009, \mn@doi [MNRAS]
  {10.1111/j.1365-2966.2009.14426.x}, 394, 1575

\bibitem[\protect\citeauthoryear{McQuinn, Zahn, Zaldarriaga, Hernquist  \&
  Furlanetto}{McQuinn et~al.}{2006}]{mcquinn-2006}
McQuinn M.,  Zahn O.,  Zaldarriaga M.,  Hernquist L.,   Furlanetto S.~R.,
  2006, ApJ, 653, 815

\bibitem[\protect\citeauthoryear{Mertens, Ghosh  \& Koopmans}{Mertens
  et~al.}{2018}]{mertens-2018}
Mertens F.~G.,  Ghosh A.,   Koopmans L.~V.~E.,  2018, \mn@doi [MNRAS]
  {10.1093/mnras/sty1207}, \href
  {http://adsabs.harvard.edu/abs/2018MNRAS.478.3640M} {478, 3640}

\bibitem[\protect\citeauthoryear{Middelberg}{Middelberg}{2006}]{pieflag-middelberg-2006}
Middelberg E.,  2006, PASA, 23, 64

\bibitem[\protect\citeauthoryear{{Morales}}{{Morales}}{2005}]{morales-2005}
{Morales} M.~F.,  2005, \mn@doi [\apj] {10.1086/426730}, \href
  {http://adsabs.harvard.edu/abs/2005ApJ...619..678M} {619, 678}

\bibitem[\protect\citeauthoryear{Morales \& Matejek}{Morales \&
  Matejek}{2009}]{morales-2009}
Morales M.~F.,  Matejek M.,  2009, \mn@doi [MNRAS]
  {10.1111/j.1365-2966.2009.15537.x}, 400, 1814

\bibitem[\protect\citeauthoryear{Morales, Hazelton, Sullivan  \&
  Beardsley}{Morales et~al.}{2012}]{morales-2012-eorwindow}
Morales M.~F.,  Hazelton B.,  Sullivan I.,   Beardsley A.,  2012, ApJ, 752, 137

\bibitem[\protect\citeauthoryear{Nuttall}{Nuttall}{1981}]{nuttall-1981}
Nuttall A.,  1981, \mn@doi [IEEE Transactions on Acoustics, Speech, and Signal
  Processing] {10.1109/TASSP.1981.1163506}, 29, 84

\bibitem[\protect\citeauthoryear{Offringa}{Offringa}{2016}]{offringa-2016-dysco}
Offringa A.~R.,  2016, \mn@doi [A\&A] {10.1051/0004-6361/201629565}, 595, A99

\bibitem[\protect\citeauthoryear{Offringa, {\TUSSEN{Bruyn}{De}{de}}~Bruyn,
  Biehl, Zaroubi, Bernardi  \& Pandey}{Offringa
  et~al.}{2010}]{offringa-2010-post-correlation-rfi-classification}
Offringa A.~R.,  {\TUSSEN{Bruyn}{De}{de}}~Bruyn A.~G.,  Biehl M.,  Zaroubi S.,
  Bernardi G.,   Pandey V.~N.,  2010, \mn@doi [MNRAS]
  {10.1111/j.1365-2966.2010.16471.x}, 405, 155

\bibitem[\protect\citeauthoryear{Offringa, {\TUSSEN{Gronde}{Van}{van}}~de
  Gronde  \& Roerdink}{Offringa
  et~al.}{2012}]{offringa-2012-scale-invariant-rank-operator}
Offringa A.~R.,  {\TUSSEN{Gronde}{Van}{van}}~de Gronde J.~J.,   Roerdink J. B.
  T.~M.,  2012, \mn@doi [A\&A] {10.1051/0004-6361/201118497}, 539

\bibitem[\protect\citeauthoryear{Offringa et~al.,}{Offringa
  et~al.}{2013}]{offringa-lofar-environment-2013}
Offringa A.~R.,  et~al., 2013, \mn@doi [A\&A] {10.1051/0004-6361/201220293},
  549, A11

\bibitem[\protect\citeauthoryear{Offringa, McKinley, Hurley-Walker
  et~al.}{Offringa et~al.}{2014}]{offringa-wsclean-2014}
Offringa A.~R.,  McKinley B.,  Hurley-Walker N.,   et~al., 2014, \mn@doi
  [MNRAS] {10.1093/mnras/stu1368}, 444, 606

\bibitem[\protect\citeauthoryear{Offringa et~al.,}{Offringa
  et~al.}{2015}]{offringa-2015-mwa-rfi}
Offringa A.~R.,  et~al., 2015, \mn@doi [PASA] {10.1017/pasa.2015.7}, 32, e008

\bibitem[\protect\citeauthoryear{Offringa et~al.,}{Offringa
  et~al.}{2016}]{offringa-2016}
Offringa A.~R.,  et~al., 2016, \mn@doi [MNRAS] {10.1093/mnras/stw310}, 458,
  1057

\bibitem[\protect\citeauthoryear{Parsons \& Backer}{Parsons \&
  Backer}{2009}]{parsons-2009}
Parsons A.~R.,  Backer D.~C.,  2009, AJ, 138, 219

\bibitem[\protect\citeauthoryear{Parsons, Pober, McQuinn, Jacobs  \&
  Aguirre}{Parsons et~al.}{2012}]{parsons-2012-arraysensitivity}
Parsons A.~R.,  Pober J.,  McQuinn M.,  Jacobs D.,   Aguirre J.,  2012, ApJ,
  753, 81

\bibitem[\protect\citeauthoryear{Patil et~al.,}{Patil
  et~al.}{2016}]{patil-2016}
Patil A.~H.,  et~al., 2016, \mn@doi [MNRAS] {10.1093/mnras/stw2277}, 463, 4317

\bibitem[\protect\citeauthoryear{Patil, Yatawatta, Koopmans  et~al.}{Patil
  et~al.}{2017}]{patil-2017}
Patil A.~H.,  Yatawatta S.,  Koopmans L. V.~E.,   et~al., 2017, \mn@doi [\apj]
  {10.3847/1538-4357/aa63e7}, \href
  {http://adsabs.harvard.edu/abs/2017ApJ...838...65P} {838, 65}

\bibitem[\protect\citeauthoryear{Peck \& Fenech}{Peck \&
  Fenech}{2013}]{peck-2013}
Peck L.~W.,  Fenech D.~M.,  2013, \mn@doi [Astronomy and Computing]
  {https://doi.org/10.1016/j.ascom.2013.09.001}, 2, 54

\bibitem[\protect\citeauthoryear{Prasad \& Chengalur}{Prasad \&
  Chengalur}{2012}]{prasad-flagcal-2012}
Prasad J.,  Chengalur J.,  2012, Experimental Astronomy, 33, 157

\bibitem[\protect\citeauthoryear{Procopio et~al.,}{Procopio
  et~al.}{2017}]{procopio-2017}
Procopio P.,  et~al., 2017, \mn@doi [PASA] {10.1017/pasa.2017.26}, 34, e033

\bibitem[\protect\citeauthoryear{Rasmussen \& Williams}{Rasmussen \&
  Williams}{2005}]{rasmussen-2005}
Rasmussen C.~E.,  Williams C. K.~I.,  2005, Gaussian Processes for Machine
  Learning (Adaptive Computation and Machine Learning).
The MIT Press

\bibitem[\protect\citeauthoryear{Scaife \& Heald}{Scaife \&
  Heald}{2012}]{scaife-heald-2012}
Scaife A. M.~M.,  Heald G.~H.,  2012, \mn@doi [MNRAS Lett.]
  {10.1111/j.1745-3933.2012.01251.x}, 423, L30

\bibitem[\protect\citeauthoryear{Trott \& Wayth}{Trott \&
  Wayth}{2017}]{trott-2017}
Trott C.~M.,  Wayth R.~B.,  2017, \mn@doi [PASA] {10.1017/pasa.2017.57}, 34,
  e061

\bibitem[\protect\citeauthoryear{Trott et~al.,}{Trott
  et~al.}{2016}]{trott-2016-chips}
Trott C.~M.,  et~al., 2016, ApJ, 818, 139

\bibitem[\protect\citeauthoryear{Vedantham, Shankar  \& Subrahmanyan}{Vedantham
  et~al.}{2012}]{vedantham-eor-foregrounds-2012}
Vedantham H.,  Shankar N.~U.,   Subrahmanyan R.,  2012, ApJ, 745, 176

\bibitem[\protect\citeauthoryear{Winkel, Kerp  \& Stanko}{Winkel
  et~al.}{2006}]{winkel-2006}
Winkel B.,  Kerp J.,   Stanko S.,  2006, AN, 88, 789

\bibitem[\protect\citeauthoryear{Yatawatta}{Yatawatta}{2014}]{yatawatta-2014-excon}
Yatawatta S.,  2014, in XXXIth URSI General Assembly and Scientific Symposium.

\bibitem[\protect\citeauthoryear{Yatawatta}{Yatawatta}{2016}]{yatawatta-co-2016}
Yatawatta S.,  2016. EURASIP (\mn@eprint {arXiv} {1605.09219})

\bibitem[\protect\citeauthoryear{Yatawatta, de Bruyn, Brentjens
  et~al.}{Yatawatta et~al.}{2013}]{yatawatta-2013}
Yatawatta S.,  de Bruyn A.~G.,  Brentjens M.~A.,   et~al., 2013, A\&A, 550

\makeatother
\end{thebibliography}
\bsp	
\label{lastpage}
\end{document}